\documentclass[largeformat]{interact} 

\usepackage{soul}
\usepackage{cancel}
\usepackage{multirow}
\usepackage{color}
\usepackage{caption}
\usepackage{makecell}
\usepackage{subcaption}
\usepackage{natbib}
 \bibpunct[, ]{(}{)}{,}{a}{}{,}%
 \def\bibfont{\small}%
 \def\bibsep{\smallskipamount}%
 \def\bibhang{24pt}%
\usepackage{graphicx}
\graphicspath{{figures/}}

\usepackage{epstopdf}
\usepackage{natbib}
\bibpunct[, ]{(}{)}{;}{a}{}{,}
\renewcommand\bibfont{\fontsize{10}{12}\selectfont}

\theoremstyle{plain}

\theoremstyle{definition}

\theoremstyle{remark}

\begin{document}
\title{Passenger Congestion Alleviation in Large Hub Airport Ground-Access System Based on Queueing Theory}

\author{
\name{Yiting~Hu\textsuperscript{a}, Xiling~Luo\textsuperscript{a,b}\thanks{CONTACT Xiling~Luo. Email: luoxiling@buaa.edu.cn} and Dongmei~Bai\textsuperscript{a}}
\affil{\textsuperscript{a}School of Electronics and Information Engineering, Beihang University, Beijing, China; \textsuperscript{b}Beihang Hangzhou Innovation Institute, Beihang University, Hangzhou, China}
}

\maketitle

\begin{abstract}
Airport public transport systems are plagued by passenger queue congestion, imposing a substandard travel experience and unexpected delays. To address this issue, this paper proposes a bi-level programming for optimizing queueing network in airport access based on passenger choice behavior. For this purpose, we derive queueing network for airport public transport system, which include the taxi, bus, and subway. Then, we propose a bi-level programming model for optimizing queueing network. The lower level subprogram is designed to correspond to the profit maximization principle for passenger transport mode choice behavior, while the upper level subprogram is designed to minimize the maximum number of passengers waiting to be served. Decision makers consider imposing queue tolls on passengers to incentivize them to change their choice and achieve the goal of avoiding congestion. Finally, we develop the successive weighted averages (MSWA) method to solve the lower subprogram's passenger share rates and the ant lion optimization (ALO) method to solve the bi-level program's queue toll scheme for upper-level objectives. We prove the effectiveness of the proposed method on two situations of simulation, daytime and evening cases. The numerical results highlight that our strategy can alleviate queue congestion for both scenarios and effectively improve evacuation efficiency.
\end{abstract}

\begin{keywords}
Queueing theory, Multinomial logit; Passenger congestion alleviation; Bi-level programming; Airport ground access system
\end{keywords}

\section{Introduction}

According to data from the \emph{ Civil Aviation Administration of China} (CAAC), the total passenger volume of the Chinese civil airports in 2018 and 2019 were 1.26 billion and 1.35 billion, increasing 10.2$\%$ and 6.5$\%$ respectively compared to the previous year. Due to the widespread COVID-19 pandemic in 2020, the passenger volume utilizing Chinese airports was only 0.86 billion, decreasing by 36.6$\%$ compared to 2019. According to the latest CAAC reports, the domestic passenger volume of the Chinese civil airports in June 2021 has exceeded the corresponding one of 2019, showing a good recovery momentum. With the global COVID-19 spread, the airport passenger flows have significantly reduced compared with the pre-pandemic period. Nevertheless, controlling the epidemic affords a sharp passenger flows recovery.

The large volume of airport passenger flows increases travel demand regarding the ground traffic around large airport hubs. As a result, in addition to the widely studied air traffic congestion \citep{pita2013integrated, lee2020dynamic}, congestion at the airport's ground transport system, involving multiple types of transport vehicles, appears more significant. According to the statistical data derived from an investigation report regarding the passengers' composition and behavior habits at the Beijing Capital International Airport (BCIA) in 2019, when passengers leave the airport, 33.1$\%$ use taxi, 16.3$\%$ the airport bus, and 15.9$\%$ the subway, while the remaining passengers use a private car or online car-hailing service. Considering private vehicles, congestion usually appears in the road network around the airport. In contrast, for public transportation such as taxis and buses, congestion appears in the road network and during the queueing process where passengers wait for available vehicles or ticket purchase. According to field research, in the taxi stand of BCIA, the most extended passenger queue length exceeds 200 passengers, and the passenger waiting time may reach 30 minutes. However, these might further increase during weekends and holidays, resulting in a poor travel experience and additional unexpected delays. Therefore, this paper focuses on optimizing the queue congestion of a public ground transport system.

The public ground transport system for passengers leaving the airport can be regarded as a typical multi-route and multistage queueing network. Formally, queueing networks are directed graphs with nodes corresponding to servers\citep{kelly1975networks}. Moreover, the service categories in the queueing network are waiting for transport, ticketing and security check. These service centers are connected according to probabilistic routing. Every service process follows the First-Come-First-Service discipline, and can be analyzed using a queueing model. Under the queueing network, the state of the airport public ground transport system can be approximated with known parameters. It can also be further optimized and combined with optimal algorithms and network performance measures (like the queue length). Additionally, the optimal design of queueing systems has been extensively studied and widely applied \citep{stolyar2005maximizing, senderovich2016conformance,zhang2020modeling}. This work adopts the queueing network perspective to examine the airport public ground transport congestion problem.

However, the routing probabilities of queueing networks corresponding to airport ground transport system is highly associated with passenger choice behavior that becomes complex due to personal characteristics and multi-objective decisions. Moreover, any changes happening in the queueing network would break its steady state of queue time and further alter passenger choices that affect the arrival in network nodes and network state. In this case, the optimization of queueing networks should be combined with the passenger transport mode choice research.

Existing studies on passenger transport mode choice in the airports are primarily based on the discrete selection model, which holds a vital assumption that passenger choice is entirely according to maximization of individual utility. In this case, the quantitative value of utility and weights corresponding to any factors for passenger choice can be approximated by mass travel data. Regarding the entire airport ground public transport system, while passengers choose transport that can maximize their individual utility, airport managers can take steps to affect passenger choice and optimize overall system performance \citep{CaoJIAOTONGUNIVERSITY}. Hence, passenger choice-making and transport system optimization can be regarded as the problems of the follower and the leader respectively, and can be further modeled utilizing bi-level programming.

For the queueing congestion problem of airport ground public transport system, this paper proposes a bi-level programming model of queueing network optimization considering passenger choice behavior. Specifically, based on queueing theory, we develop a basic airport ground public transport queueing network including three common ground access modes, under a consideration of most airport ground access situations. Under the entire queueing network, the taxi queueing sub-system using a double-ended queueing model, the sub-model of taking bus involves a classical M/M/c/K queueing model combined with a Min(N,T) renewal process, and the the subway sub-system utilizing serially connecting two M/M/c/K queueing models. The passenger arrival in queueing network is modeled by non-homogeneous Poisson process with time-vary arrival rate, and the numerical solution of state probabilities in queueing network is given. In proposed bi-level programming, decision makers impose tolls on passengers joining the queue for taking ground transports to adjust their choice behavior, optimize the queueing network and avoid queueing congestion. The lower level objective calculate passenger choice probabilities based on MNL models, and the upper level objective minimize the maximum number of passengers staying in each transport modes. We develop the method of weighted average (MSWA) to solve the lower level subprogram and the ant lion optimizer (ALO) \citep{mirjalili2015ant} to solve the upper level subprogram. Finally, to reflect the applicability of our bi-level programming model, we investigate two queue congestion scenarios of daytime and evening in the numerical simulation. The results highlight that the proposed method effectively reduces the average passenger queueing time and the number of passengers stranded in the ground-access system for both scenarios. The main contributions of this paper can be summarized as follows:

\begin{itemize}
  \item We propose bi-level programming models for airport ground public transport queueing system optimization considering passenger choice behavior. The lower level ensures passengers make choices under the principle of profit maximization based on the MNL model and successive weighted averages (MSWA) method. The upper level programming adopts the levying queue toll to adjust the passenger choice in the lower level and achieve the queueing optimization and congestion avoidance based on the Ant Lion Optimizer (ALO).
  \item We recommend a method that evaluates the state and performance of queues in airport ground public transport system considering the cases using taxi, bus, and subway based on the queueing network model. The passenger arrival in the queue model is sujected to non-homogeneous Poisson process, in which the arrival density function is obtained utilizing flight timetable. The transient solution of the queueing system state is calculated using the Runge-Kutta (RK4) method.
\end{itemize}

The airport scenario examined in this paper, as well as the underlying assumptions for the proposed model, are as follows. We consider passengers who depart the airport via ground public transport; thus, passengers arriving via airplane from other airport hubs must choose among a taxi, bus, or subway. After landing, passengers are immediately presented with the predicted waiting time and fare for each mode of transport; thus, passengers ultimately decide which mode of transports to take based on the expected wait time, fare, and other factors. From the decision-makers' perspective, the primary objective is to increase the efficiency with which passengers are escorted away and to decrease the number of passengers remaining in the queueing network. The transport fare is what decision-makers could control, and they can adjust it to influence passenger choice, with the fare difference between before and after the adjustment treated as additional queue tolls in this paper. Decision makers regularly predict the queueing network state in a day, when the predicted queue length exceeds the airport's tolerable limit, the toll program is initiated based on the proposed bi-level programming, to prevent the congestion before it happens.

The remainder of this paper is organized as follows. Section \ref{section2} presents an overview of the related research. Section \ref{section3} develops the queueing network model of the service and queueing process for the taxi, bus, and subway transportation means, and calculates their performance indicators of expected queue length and time. Section \ref{section4} introduces the passenger choice mode considered in this work., while Section \ref{section5} proposes the bi-level programming model for queueing network optimization and presents corresponding solution algorithm. Several numerical examples are presented in Section \ref{section6} demonstrating the performance of solving queuing system congestion problems. Finally, Section \ref{section7} concludes this work.

\section{Related work}\label{section2}

Scholars have made several attempts to alleviate the congestion observed in the queueing process of an airport-hub ground transport system. For the taxi queue congestion, \citet{Shi2015} uses a double-ended queueing model to simulate the taxi-taking process at large airport hubs. It discusses the proposed queueing model under the observable and unobservable cases, considering two types of customer behaviors, i.e., selfishly and socially optimal. Ref. \citet{WangFang2017} combines the customer behaviors in the double-ended queueing model with the gate policy balancing the queueing delays and operating costs. The work of \citet{Wang2019} develops a taxi dispatching algorithm exploiting the double-ended M/M/1 taxi-passenger queueing model. Current papers based on the double-ended queueing model investigate the typical taxi waiting scenario, with the taxi standing on one side of the queue and passengers at the other side. However, this paper extends current research regarding the passenger queueing model at a taxi stand by applying the appropriate changes depending on the situation.

Airport buses and the subway have similar characteristics regarding their operation, depending on a timetable involving a certain number of transport means and headway. Hence, controllable strategies such as timetable and transport rescheduling enhance the stability of the bus and subway transportation system under disruptions. The timetable rescheduling method for the sudden large-scale disruptions and daily operation is studied in \citet{cadarso2013recovery, veelenturf2016railway, zhu2020integrated, xu2020robust} and \citet{Songpo2018, Yin2019, Long2021}. Accordingly, the transport rescheduling method for potential disruptions is investigated in \citet{cadarso2013recovery, kroon2015rescheduling}. In \citet{liebchen2008first, lu2016optimization, chen2017design} the authors focus on timetable and route design for long-term decision-making. All the researches mentioned above focus on the process once the transport departs the platform, while in this study, we consider the passenger queueing process before departure, which is a new point of view.

\section{Queueing network of airport ground public transport system}\label{section3}

The airport ground transport system comprises various airport ground-access modes. In this work, we consider three main public transport modes, taxi, airport bus and subway in airport to found queueing network and study its optimization.

The queueing network's arrival process describes the behavior of passengers upon their arrival \citep{herzog2000formal}. While the flight schedule primarily determines when passengers arrive at the airport, it is difficult to determine when passengers arrive at the location where public transportation is available. Before leaving the airport, passengers may perform various behaviors according to personal demands, e.g., claiming baggage, using the bathroom, and shopping. As a result, passengers remain at the airport for a variable amount of time after disembarking, and the passenger arrival process at public transportation becomes stochastic. The Poisson distribution is the most frequently used probabilistic model for customer arrivals \citep{sayarshad2016survey}, such as passenger arrival at an airport security check \citep{wang2017application} or transit bus stop \citep{cai2018stochastic}. Hence, the Poisson distribution can also be used to simulate passenger arrivals at airport public transportation. Additionally, there is a strong correlation between the rate of passenger arrivals at ground transportation services and the daily schedule. As a result, it makes more sense to model passenger arrival using the non-homogeneous Poisson process. Since this article does not examine specific passenger arrival models in public transportation, we directly average the passenger data in-flight to obtain the arrival density function for non-homogeneous Poisson passenger arrivals in public transportation.

\begin{figure}
\centering
\includegraphics[width=6in]{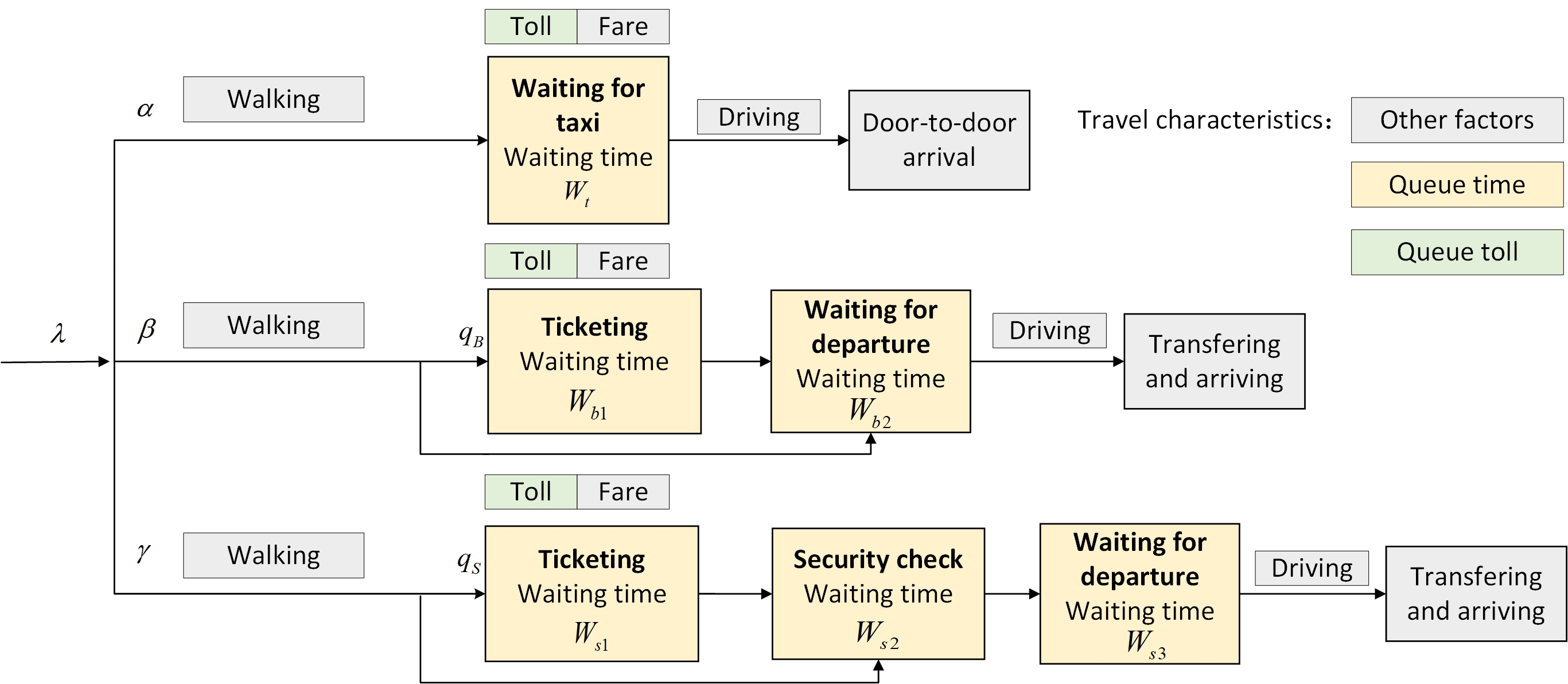}
\caption{Queueing network that depicts the process of taking ground public transports.\label{hu1}}
\end{figure}

The passengers who depart the airport choose a particular traveling mode of taxi, bus and subway, and join the corresponding queueing system following a non-homogeneous Poisson process of rate $\lambda(t)$. Each access mode is subjected to the first-come-first-serve discipline. The percentages of passengers choosing a taxi, airport bus, and subway are $\alpha(t)$, $\beta(t)$, and $\gamma(t)$, respectively, which satisfy
\begin{equation}\label{eq1}
\alpha(t)+\beta(t)+\gamma(t)=1
\end{equation}
at time $t$, thus the passenger arrival rates in the corresponding queueing systems are $\lambda_{X}(t)=\alpha(t)\lambda(t)$, $\lambda_{B}(t)=\beta(t)\lambda(t)$ and $\lambda_{S}(t)=\gamma(t)\lambda(t)$, respectively. The entire procedure of taking ground public transport to depart, and travel characteristics during this during process are illustrated in Fig. \ref{hu1}, in which the yellow blocks represent travel characteristics corresponding to queue time, the green block represents additional queue tolls imposed on passengers and the grey blocks are all the other retained characteristics. It demonstrates that passengers need only to wait for a taxi to arrive; to take the airport bus, passengers purchase tickets at the ticket office with probability $q_B$ (passengers purchase online tickets in advance with probability $1-q_B$) and wait for the bus to depart; and to take the subway, passengers purchase tickets on the spot with probability $q_S$ (passengers employ subway
cards or QR codes with probability $1-q_S$), pass through security, and wait for the train to arrive. Below, we model and solve the service processes associated with taxis, buses, and the subway.

The performance indicator for queueing networks is primarily concerned with the number of stranded crowds and the length of time spent waiting. The waiting time is spent standing still, which includes queuing for taxis, tickets, and security checks, as well as waiting for bus and train departures. Hence, we model the above services as service centers for the airport ground access system queueing network in this section. In the bi-level model described in Section \ref{section5}, passengers should be informed of each transport's transient predicted waiting time; thus, we focus on the queueing network's transient state, for which the initial condition of the current queueing network state is required.

\subsection{Queueing Model For Taxi}\label{section3.1}
Passengers feeling tired after a long journey and needing a door-to-door service always utilize taxis as the first choice to arrive at their destination directly, quickly, and comfortably. The service process of a taxi can be described as follows. Taxis queue up in the taxi queueing pool for passengers, while passengers queue up in the waiting area to get in a taxi, forming a typical double-ended queueing system. The taxi service process is shown in Fig. \ref{hu2}. Both taxi arrival and passenger arrival follow the Poisson distribution in this work.

When the matching time between passengers and taxis is relatively short compared with the inter-arrival time of passengers and taxis, it can be negligible\citep{Shi2015}. However, to maintain order, some airports control the matching process. For instance, at the Beijing Capital International Airport, taxis need the dispatcher to release them from the taxi queueing pool after arriving at the airport and before carrying passengers. In this way, order management can be realized at the cost of increasing the matching time between taxis and passengers, which can no longer be negligible. Further more, even if the airport does not control the matching between taxis and passengers, still, the matching time cannot be made negligible given the short inter-arrival time of taxis and passengers at peak time. Thus, the taxi taking process is modeled by the double-ended queueing system with nonzero matching time in this paper, which has been studied by \citep{shi2015study}.

We assume that the taxi arrival process is subjected to non-homogeneous Poisson distribution with time-varying rate $\lambda_T(t)$, and the matching time between taxis and passengers follows exponential distribution with constant rate $\mu$. The maximum number of taxis allowed joining the queue for passengers is $K_{T}$, and waiting room for passengers is unlimited. The taxi-passenger system is modeled as a two-dimensional Markov process \{$I_1(t),I_2(t),t\geq0$\}, where $I_1(t)$ and $I_2(t)$ represent the number of passengers and drivers in the system at time $t$ respectively. The state space of Markov process is $S_2 = \{(i,j), i\in \{0,1,2,...\}, j\in \{0,1,2,...,K_{T}\}\}$. Defining the level $n$ as number of passengers in the system and utilizing the level-dependent method \citep{lian2008tandem} to arrange the states as
\begin{center}
Level 0: (0,0), (0,1), (0,2),...,(0,$K_T$);\\
Level 1: (1,0), (1,1), (1,2),...,(1,$K_T$);\\
  $\vdots \qquad \qquad\qquad \vdots$\\
\end{center}

The infinitesimal generator matrix of passenger-taxi Markov process $Q_X(t)$ is
\begin{equation}\label{eq2}
Q_X(t)=
\left[
  \begin{array}{ccccc}
B_0(t)& A(t)& & \quad& \quad \\
C & B_1(t) & A(t) & &   \\
      & C & B_1(t) & A(t) &  \\
      &  &\ddots & \ddots &  \ddots \\
  \end{array}
\right],
\end{equation}
in which blocks $A(t)$, $B_0(t)$, $B_1(t)$, $C\in R^{(N+1)\times(N+1)}$ are transition matrices of the states transiting from level determined by row number to level determined by column number. Therefore,
\begin{equation}
B_0(t)=
\left[
  \begin{array}{ccccc}
-(\lambda_T(t)+\lambda_X(t)) &\lambda_T(t)&  &  &   \\
 & -(\lambda_T(t)+\lambda_X(t)) &\lambda_T(t) & &   \\
 &  &\ddots & \ddots &  \ddots \\
  &   &   & -(\lambda_T(t)+\lambda_X(t)) &\lambda_T(t) \\
 & & & &-\lambda_X(t)\nonumber\\
  \end{array}
\right],
\end{equation}
\begin{equation}
B_1(t)=
\left[
  \begin{array}{ccccc}
-(\lambda_X(t)+\mu) &\lambda_T(t)&  &  &   \\
 & -(\lambda_T(t)+\lambda_X(t)+\mu) &\lambda_T(t) & &   \\
 &  &\ddots & \ddots &  \ddots \\
  &   &   & -(\lambda_T(t)+\lambda_X(t)+\mu) &\lambda_T(t) \\
 & & & &-\lambda_X(t)+\mu\nonumber\\
  \end{array}
\right],
\end{equation}
\begin{equation}
A(t)=diag\left[\lambda_X(t), \lambda_X(t),...,\lambda_X(t)\right],\nonumber
\end{equation}
\begin{equation}
C=
\left[
  \begin{array}{cc}
 O_{1\times (K_T-1)} &  0 \\
  \mu I_{K_T\times K_T}&  O_{(K_T-1)\times 1} \nonumber\\
  \end{array}
\right].
\end{equation}

Define $p_{i}(t) = [p_{i,0}(t), p_{i,1}(t),...,p_{i,K_T}(t)](i\leq 1)$ as the probability that there are $i$ passengers and $j(1\leq j\leq K_T)$ taxis in the system at time $t$, and the state probability vector of system can be written as $P_X(t)=[p_{1}(t), p_{2}(t),...]$. Based on the Kolomogorov's Backward Equations, we have
\begin{equation}\label{eq3}
P_X^{'}(t)=P_X(t)Q_X(t).
\end{equation}

Similar to \citet{Tirdad2016}, we employ the fourth-order Runge-Kutta method (RK4) to calculate the transient state probability $P_X(t)$. By defining the step size as ${\Delta}t$, we discrete continual $t$ and $P_X(t)$ as $t_{n}=n{\Delta}t$ and $P_{X}(t_{n})$. The RK4 method provides the $\kappa_{1}, \kappa_{2}, \kappa_{3}$ and $\kappa_{4}$ to obtain $P_X(t_{n+1})$ from given $P_X(t_{n})$.
\setlength{\arraycolsep}{0.0em}
\begin{eqnarray}
\kappa_{1}&{}={}&Q_X(t_{n})P_X(t_{n})\Delta t,\\
\kappa_{2}&{}={}&Q_X(t_{n}+\frac{{\Delta}t}{2})(P_X(t_{n})+\frac{\kappa_{1}}{2})\Delta t,\\
\kappa_{3}&{}={}&Q_X(t_{n}+\frac{{\Delta}t}{2})(P_X(t_{n})+\frac{\kappa_{2}}{2})\Delta t,\\
\kappa_{4}&{}={}&Q_X(t_{n}+\Delta t)(P_X(t_{n})+\kappa_{3})\Delta t,\\
P_X(t_{n+1})&{}={}&P_X(t_{n})+\frac{1}{6}(\kappa_{1}+2\kappa_{2}+2\kappa_{3}+\kappa_{4}).
\end{eqnarray}

For the order of $Q_X(t)$ in numerical implementation, we choose a sufficiently large integer number such that the probability of have corresponding passengers in the queue is negligible, as the order of $Q_X(t)$. For a given initial condition, the transient state probability at every ${\Delta}t$ can be obtained by RK4 method. The expected queue length of passengers $L_X(t_n)$ is easily calculated from state probability. Exploiting Little's law, the expected passenger sojourn time is
\begin{equation}\label{eq9}
W_{X}(t_{n})=\frac{L_{X}(t_{n})}{\lambda_{X}(t_{n})}.
\end{equation}

\subsection{Queueing Model For an Airport Bus}\label{section3.2}

The airport bus service system can be divided into two processes, i.e., buying tickets and waiting for departure. The bus can transport multiple passengers simultaneously and has an appealing riding comfort, but it does not directly go to the passenger's destination, and passengers usually spend some time at the waiting area. Regarding the airport bus service system, passengers should first buy tickets and then get on the bus. Some passengers may buy tickets at the ticket office with probability $q_{B}$, and the remaining ones buy e-tickets in advance with probability $1-q_{B}$. The ticket office has $c_{B}$ ticket counters that open for ticket sales at the same time. We assume that passengers immediately start with the waiting process to ultimately departure after getting tickets and treat other procedures for taking bus (ticket check and placing luggage) as part of the departure waiting process. These two steps will be modeled respectively.

The queueing process in the buying tickets step is treated as an M/M/c/K model where the passenger arrival rate is $\lambda_{B1}(t)=q_{B}\lambda_{B}(t)=q_{B}\beta(t)\lambda(t)$. The passenger input is a Poisson process with parameter $\lambda_{B1}(t)$ and the time passengers spend buying tickets is subject to a negative exponential distribution with parameter $\mu_{B}$, which is the ticket sale rate per ticket counter. The queueing model utilization of this service process is $\rho_{B}=\frac{\lambda_{B1}}{c_{B}\mu_{B}}$. The waiting room for passengers is unlimited.

We define the state probability in ticketing system as $P_B(t)=[p_{B,0}(t),p_{B,1}(t),p_{B,2}(t),...]$, where $p_{B,j}(t)$ is the probability that $j$ passengers are waiting for ticketing at time $t$. Transition matrix for M/M/c/K ticketing process $Q_B(t)$ and system dynamic equation are\begin{eqnarray}\label{eq10}
&&Q_{B}(t)=\nonumber\\
&&\left[
  \begin{array}{cccccc}
    -\lambda_{B1}(t)& \lambda_{B1}(t)& & & & \\
    \mu_{B} & -(\mu_{B}+\lambda_{B1}(t)) & \lambda_{B1}(t) & & &\\
     & 2\mu_{B} & -(2\mu_{B}+\lambda_{B1}(t)) & \lambda_{B1}(t)  & &\\
     & &\ddots & \ddots &  \ddots & \\
     &  &  & c_{B}\mu_{B} & -(c_{B}\mu_{B}+\lambda_{B1}(t)) & \lambda_{B1}(t)\\
     &  &  &  & c_{B}\mu_{B} & -c_{B}\mu_{B}\\
  \end{array}
\right],
\end{eqnarray}
\begin{equation}
P_B^{'}(t)=P_B(t)Q_B(t).
\end{equation}
where the order of $Q_B(t)$ is approximated as $K_B$ in the same way as $Q_X(t)$ introduced in Section \ref{section3.1}

The transient solution of $P_B(t)$ can be computed utilizing RK4, as presented Section 3.1. The number of expected sojourn passengers $L_{B1}(t)$ and the corresponding waiting time $W_{B1}(t)$ are
\begin{equation}\label{eq11}
L_{B1}(t) = P_B(t)[0,1,2,...,K_B]^T,
\end{equation}
\begin{equation}\label{eq13}
W_{B1}(t) = \frac{L_B(t)}{\lambda_B(t)}.
\end{equation}

The passenger arrival in the bus waiting process is considered as a Poisson distribution with parameter $\lambda_{B2}(t)$. this passenger arrival comes from passengers walking out of the ticket office and passengers buying tickets online. Hence, $\lambda_{B2}(t)$ is expressed as
\begin{equation}
{\lambda}_{B2}(t)=\lambda_{B}(1-q_{B})+min(\lambda_{B1},\mu_{B})).
\end{equation}

The bus departure schedule follows the Min$(N,T)$-policy, that is, the bus starts either when it is full of $N$ passengers or time $T$ has passed after the last bus has departed. According to the previous analysis, the bus departure process is regarded as a renewal process, and the update interval is a distribution of Min$(N,T)$, where $N$ is the bus capacity and $T$ a fixed time. 
Let the $m^{th}$ bus departure time in the renewal process be $s_{m}$ and passengers waiting for departure are cleared at $s_{m}$. The bus service system enters an idle period with Min$(N,T)$ renewal policy after every clearance, and the renewable time $s_{m}$ is an embedded Markov chain \citep*{Deng1998}. We define the time from the last renewable time $s_{m-1}$ until the $n^{th}$ passenger enters the system as $t^{n}$, and the distribution of $t^n$ as $F_{n}(t)$. The input is a Poisson process with parameter $\lambda_{B2}(t)$, hence the passenger arrival time interval sequence $\left\{t^{n}-t^{n-1}\right\}(0<n\leq N)$ approximately subjects to a negative exponential distribution with approximate parameter $\overline{\lambda_{B2}}(t)$. Therefore, $F_{n}(t)$ is an $n$-order Erlang distribution. Define the initial time in a renewal process as $t_0$, and the probability density $f_{n}(t)$ and the distribution function $F_{n}(t)$ of residual life in current renewal interval are written as
\setlength{\arraycolsep}{0.0em}
\begin{eqnarray}
f_{n}(t)&{}={}&\frac{\overline{\lambda_{B2}}(t)^{n}t^{n-1}}{(n-1)!}e^{-\overline{\lambda_{B2}}(t)t}, 0<t\leq T-t_0,\label{eq15}\\
F_{n}(t)&{}={}&1-\sum_{j=0}^{n-1}\frac{\overline{\lambda_{B2}}(t)^{j}}{j!}e^{-\overline{\lambda_{B2}}(t)t},0<t\leq T-t_0.\label{eq16}
\end{eqnarray}

We further define the initial condition of current number of passengers waiting on the bus at $t_0$ as $m_0$, thus if the accumulated passengers arriving in the residual time of this renewal interval reach $N-m_0$, the bus would depart advance. Therefore, the passengers' probability of getting a service, as calculated by the $T$-policy and the $N$-policy, are $F_{N-m_0}(T-t_0)$ and $1-F_{N-m_0}(T-t_0)$, respectively, and the corresponding expected passenger waiting time is $W_{B2}^{T}$ and $W_{B2}^{N}$:
\begin{equation}\label{eq17}
W_{B2}^{T}=T-t_0,
\end{equation}
\begin{equation}\label{eq18}
W_{B2}^{N}=\frac{E[t^{2}]}{2E[t]} =\frac{N-m_0+1}{2\overline{\lambda_{B2}}(T-t_0)},
\end{equation}
where $E[t]$ and $E[t^{2}]$ are the first and the second moment of distribution $F_{N-m_0}(t)$ \citep{Deng1998}. According to the above formulas, the mean passenger waiting time under the Min$(N,T)$-policy is
\begin{eqnarray}\label{eq19}
W_{B2}&{}={}&W_{B2}^{T}F_{N-m_0}(T-t_0)+W_{B2}^{N}(1-F_{N-m_0}(T-t_0))\nonumber\\
&{}={}&\frac{N+1}{2\lambda_{B2}}F_{N-m_0}(T-t_0)+\frac{T}{2}(1-F_{N-m_0}(T-t_0)).
\end{eqnarray}

The expected passenger sojourn time in the airport bus queueing system $W_{B}(t)$ is the sum of the passenger waiting time in the ticket and renewal process
\begin{eqnarray}\label{eq20}
W_{B}(t)&{}={}&\frac{L_{B}(t)}{{\lambda}_{B1}(t)}+\frac{T}{2}(1-F_{N-m_0}(T-t_0))\nonumber\\
&&{+}\:\frac{N+1}{2\lambda_{B2}}F_{N-m_0}(T-t_0).
\end{eqnarray}
\subsection{Queueing Model For Subway}

\begin{figure}
        \centering
		\begin{subfigure}{.45\textwidth}
            \includegraphics[width=\textwidth]{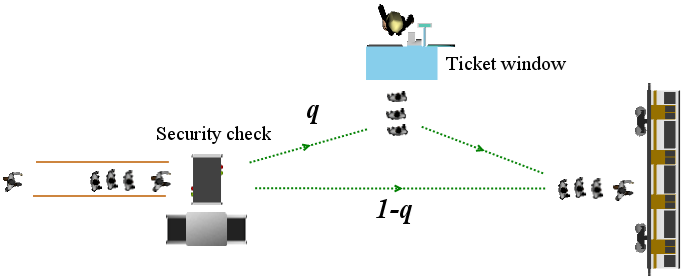}
			\caption{Ticketing after security check}
		\end{subfigure}
		\begin{subfigure}{.45\textwidth}
            \includegraphics[width=\textwidth]{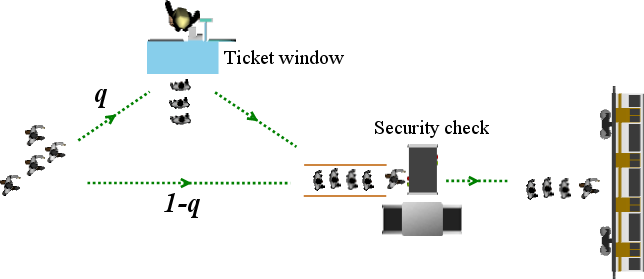}
			\caption{Ticketing before security check}
		\end{subfigure}
        \caption{Subway service process.}
        \label{Hu3}
	\end{figure}

Utilizing the subway affords the lowest cost among all three access modes discussed in this paper. However, its accessibility and comfort are relatively low. The subway service system comprises three steps: passing the security check, buying tickets and waiting for the train to depart. Depending on the subway station, the order of passing the security check and buying tickets may be reversed (Fig. \ref{Hu3}). We define the passenger arrival rates at the security check as $\lambda_{S1}$ and for buying tickets as $\lambda_{S2}$. Passengers queueing up to buy tickets have a probability of $q_{S}$, and the remaining ones employing subway cards or QR codes directly proceed to the next step with a probability of $1-q_{S}$. Passing the security check and buying tickets are regarded as negative exponential distributions with parameters $\mu_{S1}$ and $\mu_{S2}$, respectively, and the fixed number of security check servers and ticket counters are $c_{S1}$ and $c_{S2}$, respectively. The security check and ticket queueing models constitute a M/M/c-M/M/c tandem service system, with the queueing model utilization for passing the security check and buying tickets being $\rho_{S1}=\frac{\lambda_{S1}}{c_{S1}\mu_{S1}}$ and $\rho_{S2}=\frac{\lambda_{S2}}{c_{S2}\mu_{S2}}$, respectively. The expected waiting time for security check and ticket purchase are $W_{S1}(t)$ and $W_{S2}(t)$, the train headway has a fixed time $M$, and the passenger sojourn time in the entire subway service system is $W_{S}$.

For the subway queueing model, the order of passing the security check and buying tickets may be reversed depending on the subway station. If the security check is the first process, the passenger arrival rate is $\lambda_{S1}(t)=\lambda_{S}(t)$. If $\lambda_{S1}(t)$ is less than the overall service rate $c_{S1}\mu_{S1}$, the passenger output rate is the same as the input. Otherwise, the passenger output rate is equal to the service rate, and thus the output is a Poisson process of $c_{S1}\mu_{S1}$. Furthermore, every passenger has a probability $q_{S}$ to queue up for tickets and the passenger arrival rate for tickets is $\lambda_{S2}(t)=q_{S}\min(c_{S1}\mu_{S1},\lambda_{S}(t))$. If buying tickets is the first process, passengers without any prepaid card, month-card, or e-card should first buy a one-way ticket at the ticket counter with a probability $q_{S}$. Hence, $\lambda_{S2}(t)=q_{S}\lambda_{S}(t)$ and $\lambda_{S1}(t)=\min(\lambda_{S2}(t),c_{S2}\mu_{S2})+(1-q_{S})\lambda_{S}(t)$.

The sojourn passengers in the security check and the ticket system are independent, and we also choose two sufficiently large numbers to approximate their transition matrix orders. Similar as above sections, this strategy affords to compute the transient solutions of these systems utilizing RK4.

The passenger sojourn time in the overall subway service system is
\begin{equation}\label{eq21}
W_{S}(t)=\frac{L_{S1}(t)}{\lambda_{S}}+q_{S}\frac{L_{S2}(t)}{\lambda_{S2}(t)}+M,
\end{equation}
where $M$ refer to the time of train departure away from now.

For the sake of simplicity, this paper considers during simulations the security check first case. Given that in various countries, some subway stations do not have a security check process, the passenger sojourn time is obtained by neglecting $L_{S2}$, which is beyond the scope of this paper.

The two queueing network performance criteria reflecting the congestion intensity of the overall ground access system are
\begin{eqnarray}
  W&{}={}&{\alpha}W_{X}+{\beta}W_{B}+{\gamma}W_{S},\label{eq22}\\
  L&{}={}&\max\{L_t,L_b,L_s\},\label{eq23}
\end{eqnarray}
where $W$ and $L$ represents the mean passenger sojourn time and the maximum stranded crowd number among three transport modes.

\section{Passenger transport mode choice}\label{section4}
When choosing the available public transport mode, passengers have a comprehensive consideration of the utility they would gain from taking different transport modes and decide according to the maximization of their utility. They usually take into account factors that include cost, travel time, accessibility, comfort, among many others. Nevertheless, the weights and converted values of such factors differ for each passenger due to their personal characteristics and preferences. Passengers with similar individual features are more likely to hold the same evaluation standards. Hence, when modeling passenger choice behavior, passengers are classified into several categories to fit specific models, respectively.

This paper utilizes MNL model to model passenger choice among taxis, airport buses and subways. It assumes that passengers are classified into $H$ categories, and the probability of the $h^{th}(0<h\leq H)$ class of passenger choosing the $i^{th}$ travelling option from the set of taxis, airport buses and subways is
\begin{eqnarray}
P_{h}(i)&{}={}&P(U_{ih}\geq{U_{jh}},{\forall}j\in \{X,B,J\},i\neq j),\\
U_{ih}&{}={}&V_{ih}+\epsilon_{ih},
\end{eqnarray}
where $U_{ih}$ is the utility of the transport mode $i$, $V_{ih}$ is the measurable part and $\epsilon_{ih}$ is the random part of utility.

For each category of passengers, some factors contributing to transport utility, such as accessability, fare, and in-vehicle time, can be determined and remain unchanged since they have decided their destination. Conversely, the utility corresponding to the queue time for taking transport can not be determined because the waiting time in ever-changing queueing system is hard to predict for passengers. Besides, queue tolls imposed by decision makers are also a dynamic character in a passenger's consideration. In this context, this paper studies ground transport from the queueing perspective. It simply views utility, apart from queue time and tolls, as a total static part, and divides the transport mode utility into three parts: utility of queue time ($W_{ih}$), utility of queue tolls ($J_{ih}$) and utility of other factors that stay constant ($O_{ih}$). Thus, we have

\begin{equation}\label{eq26}
V_{ih}=W_{ih}\omega_{ih}^{T}+O_{ih}\omega_{ih}^{O},+J_{ih}\omega_{ih}^{J}
\end{equation}
where $\omega_{ih}^{W}$, $\omega_{ih}^{O}$ and $\omega_{ih}^{J}$ are weights of $W_{ih}$, $O_{ih}$ and $J_{ih}$. In the subscripts of the above notations, $i$ and $n$ represent the $i^{th}$ transport mode and the $h^{th}$ class of passengers. Based on the assumption of the gumbal distribution of $\epsilon_{ih}$, for the given $V_{ih}$, the probability $P_{h}(i)$ can be rewritten as
\begin{equation}\label{eq27}
P_{h}(i)=\frac{e^{{V_{ih}}}}{\sum_{i\in \{X,B,J\}}e^{{V_{ih}}}}.
\end{equation}

In practice, when determining the passenger choice model parameters, the sample information can be obtained through interviews, questionnaires and databases. Additionally, the maximum likelihood estimation method can calibrate the model parameters.

\section{Bi-level programming model of queue length optimization}\label{section5}

We employ Bi-level programming to simulate passenger choice and optimize airport ground public transport queueing network, in which the lower level function describes the passenger choice behavior, and upper level programming optimizes queueing network by levying tolls.

The optimization model of passenger transport mode choice is expressed as
\begin{eqnarray}\label{eq28}
&&\min \quad U=\sum_{i\in \{X,B,J\}}\sum_{h\in H}\int_o^{P_{ih}\lambda}D^{-1}(x)dx\nonumber\\
&&s.t.\quad \sum_{i\in \{X,B,J\}}\sum_{h\in H} P_{ih}=1.
\end{eqnarray}
where $U$ is the total utility obtained by a passenger through a mode of transport to leave the airport. Then, $D^{-1}(P_{ih}\lambda)$ is the generalized cost function of passenger flow in transport $i$. The objective function Eq. \ref{eq28} means that the passenger behavior follows the principle of maximization of individual utility.

From the airport manager perspective, they expect to avoid extreme congestion occurring at any service center, and hope for the highest efficiency in the transport system for the passengers. Thus, the upper level function is
\begin{eqnarray}\label{eq29}
\min \quad W =\max &&\{L_X,L_B, L_S\},
\end{eqnarray}
where $L_X$, $L_B$ and $L_S$ are number of passengers waiting in the service centers of the queueing network.
\subsection{Method of weighted average for lower level subprogram}\label{section5.1}
The airport ground public transport choice problem examined in lower level subprogram is based on MNL model presented in Section \ref{section4}. For the given utility of queue tolls $J_{ih}$, static utility $O_{ih}$ and all the weights off utility ($\theta_{ih}$), there are a feedback between $T_{ih}$ and passenger share rate $\alpha$, $\beta$ and $\gamma$. That is, any perturbation for passenger share rate would cause changes in $T_{ih}$ according to queueing model, and which would also redistribute passenger share rate among transport modes according to MNL model, until to equilibrium. Based on this analysis, we utilize the Method of weighted average to solve the lower level subprogram of passenger choice.

\begin{figure}
\centering\includegraphics[width=2.2in]{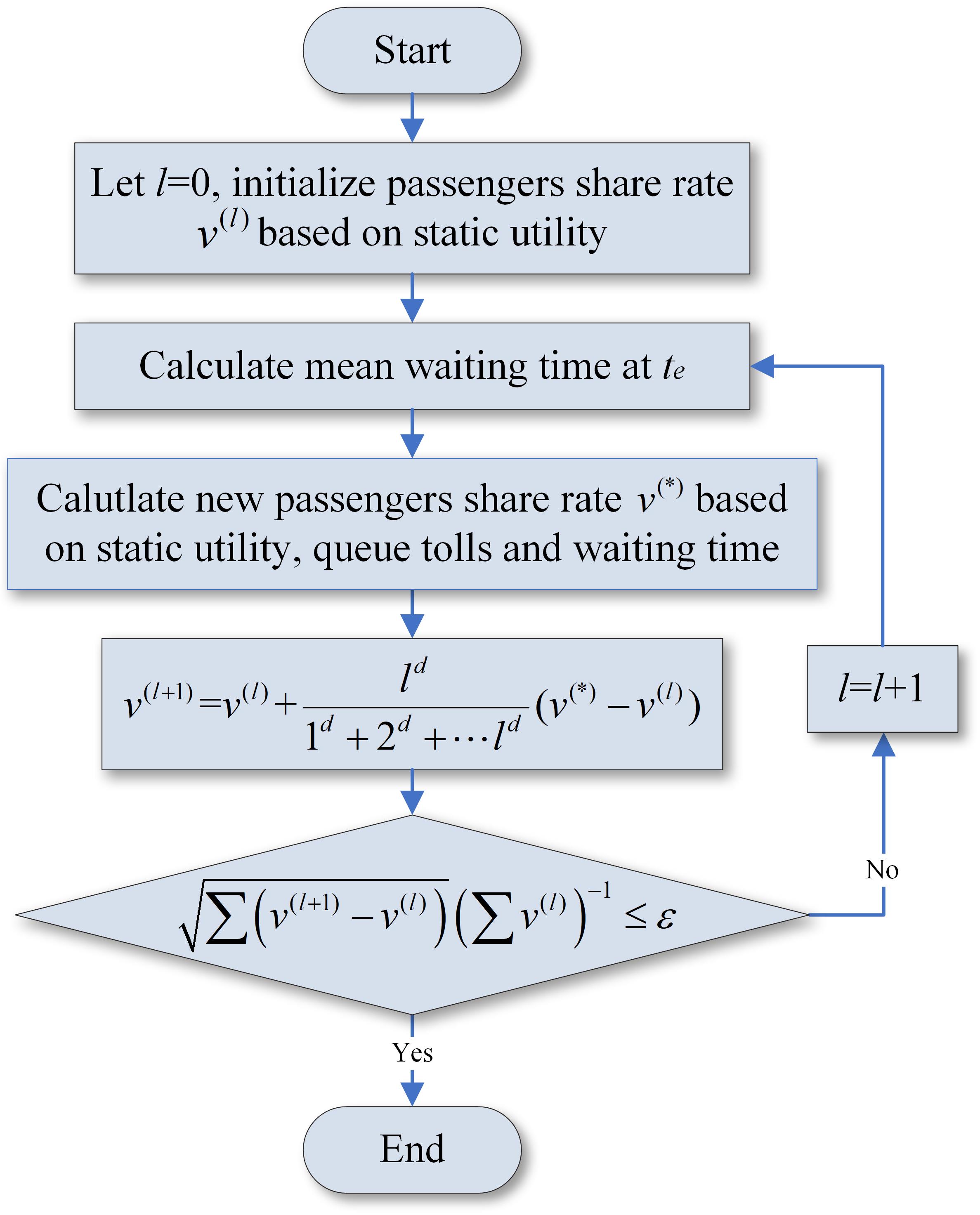}
\caption{Flow chart of the MSWA iteration.}
\label{Hu4}
\end{figure}

The entire iteration process is modeled utilizing the MSWA method illustrated in Fig. \ref{Hu4}, which is summarized as follows:
\begin{enumerate}
\item Iteration starts from $l$=0 and the initial condition is that all queue time $W_{ih}$ are zero. Thus the initial passenger share rate among taxi, bus and subway is calculated by Eq. \ref{eq28} as $v^{(0)}=[\alpha^{(0)}=P(X),\beta^{(0)}=P(B),\gamma^{(0)}=P(S)]$.
\item Calculate the predicted mean queueing time $W_{X}(t)$, $W_{B}(t)$ and $W_{S}(t)$ at time $t$ with the passenger share rate in current iteration $v^{(l)}=[\alpha^{(l)},\beta^{(l)},\gamma^{(l)}]$. By combining the static utility and queue tolls with the dynamic queueing time, we obtain $V_{X}=\omega^{T}_XW_{X}(t_{e})+\omega^{O}_XO_{X}+\omega^{J}_XJ_{X},  V_{B}=\omega^{T}_BW_{B}(t_{e})+\omega^{O}_BO_{B}+\omega^{J}_BJ_{B}$, and $V_{S}=\omega^{T}_SW_{S}(t_{e})+\omega^{O}_SO_{S}+\omega^{J}_SJ_{S}$. Then the intermediate passenger share rate $v^{(*)}=[\alpha^{(*)},\beta^{(*)},\gamma^{(*)}]$ is calculated exploiting the MNL model.
\item The next passenger share rate $v^{(l+1)}$ is updated by
\begin{align}
&v^{(l+1)}=v^{(l)}+\chi^{(l)}(v^{(*)}-v^{(l)}),\\
&\chi^{(l)}=\frac{l^{d}}{1^{d}+2^{d}+3^{d}+\cdots+l^{d}},
\end{align}
where $\chi^{(l)}$ is the step size in the $l^{th}$ iteration and $d$ is a given constant. In our trials we set $d=1$.
\item If the error between $v^{(l + 1)}$ and $v^{(l)}$ in last iteration satisfies
    \begin{equation}
     \left\|v^{(l + 1)}-v^{(l)}\right\|_2\left\|v^{(l)}\right\|_1^{-1}\leq\epsilon,
    \end{equation}
where $\epsilon$ is the given threshold, we deem that the iteration converges. Hence, the passenger share rates and queueing time reach an equilibrium, and our method provides $v^{(l+1)}$ as the final solution. Otherwise, iteration returns to Step 2, sets $n=n+1$ and re-executes the loop.
\end{enumerate}

\subsection{ALO algorithm for upper lever subprogram}\label{section5.2}

For every group of given queue tolls, a corresponding passenger choice can be solved in the lower lever subprogram. In upper subprogram, we utilize Ant Lion Optimizer (ALO) to generate the queue toll scheme to minimize Eq. \ref{eq29}.

The ALO is a novel nature-inspired evolutionary algorithm which mimics the interaction between antlions and ants as antlions hunting ants. In this interaction, the ants move over the search space for food with random walk movement at every step of optimization and the fitness (objective) function is utilized to mark ant positions. The antlions are also random located in the search space with a fitness. To hunt ants, antlions build a trap around themselves whose radii are proportional to their fitness for limiting ants' movement. After several iterations, ants slides to the bottom of traps and are caught by antlions, and antlions optimize its fitness by their catch of ants.

We set the price of queue tolls of three transport modes as a three dimension search space in ALO. The lower and upper limits of every transport queue toll are $Cl$ and $Cu$ respectively. The maximum iterations is $t^{max}$. The fitness of every ant and antlion is calculated using Eq. \ref{eq29} and Eq. \ref{eq28}, and the more little the Eq. \ref{eq29}, the fitter the ant/antlion.

At every iteration $t$, the random walk is generated with $r(t)$ which is either -1 or 1, thus the position vector of ant during the iteration is $X(t)=[0,cumsum(r(t_1)),cumsum(r(t_2)),...,cumsum(r(t^{max}))]$ where cumsum is the cumulative sum function. With given position vector $X_i(t)$ of the $i^{th}$ ant, we still need a normalization to bound the random walk in the search space at every iteration $t$, hence
 \begin{equation}\label{eq25}
   X^t_i=\frac{(X^t_i-\min X_i(t))\times(Cu_{i}^t-Cl_{i}^t)}{\max X_i(t)-\min X_i(t)}+Cl^t_i,
 \end{equation}
where $Cu_{i}^t$ and $Cl_{i}^t$ are upper and lower bound for the $i^{th}$ and in the $t^{th}$ iteration.

The complete algorithm of ALO are presented as follows:
\begin{enumerate}
\item Initialize the first population of ants and antlions in the search space randomly
    \begin{eqnarray}
     a^0_{i,k}=Cl+rd[0,1]\times(Cu-Cl), k\in{X,B,S},\\
     x^0_{i,k}=Cl+rd[0,1]\times(Cu-Cl), k\in{X,B,S},\\
    \end{eqnarray}
where $a^0_{i,k}$ and $x^0_{i,k}$ are the $i^{th}$ ant and antlion, and the subscript $k$ denotes the corresponding dimension. $rd[0,1]$ generate a random value between 0 and 1.
\item Calculate their fitness using Eq. \ref{eq29} under the corresponding passenger share rate solved by MSWA and determine the elite antlion $x^*=[x^*_X,x^*_B,x^*_S]$ with the best fitness.
\item For each ant $i$, select an antlion $x_{i}^{t}$ to update the upper and lower limits of ant's value bounded by this antlion:
    \begin{eqnarray}
     Cl^{t+1}_{i}=x_{i}^{t}+sign(rd[-1,1])\times\frac{Cl^{t+1}_{i}}{10^{w^t}\times(t/t_{max})}, \\
     Cu^{t+1}_{i}=x_{i}^{t}+sign(rd[-1,1])\times\frac{Cu^{t+1}_{i}}{10^{w^t}\times(t/t_{max})},\\
    \end{eqnarray}
    where $w^t$ controls the rate of ants sliding to antlions, that is the reduction rate of trap radii. Create the random walk of ant $i$ bounded by antlion $x_{i}^{t}$ using Roulette wheel as $R^t_A$. Besides, ant $i$ also generate a random walk bound by elite antlion $x^*$ with $R^t_E$, and the real random walk in the iteration $t+1$ is
    \begin{equation}\label{}
      a^{t+1}_i=\frac{R^t_A+R^t_E}{2}
    \end{equation}
\item For each ant $i$, if it is fitter than its corresponding antlion $i$, replace the antlion with it:
    \begin{equation}\label{}
      x^{t+1}_i=a^{t+1}_i.
    \end{equation}
\item Update the elite of antlions in the current iteration.
\item If the number of iteration $t$ exceeds the max iteration $t^{max}$, end the algorithm. Otherwise iteration returns to step 3 and re-executes the loop.
\end{enumerate}

\begin{figure}
\centering\includegraphics[width=4.2in]{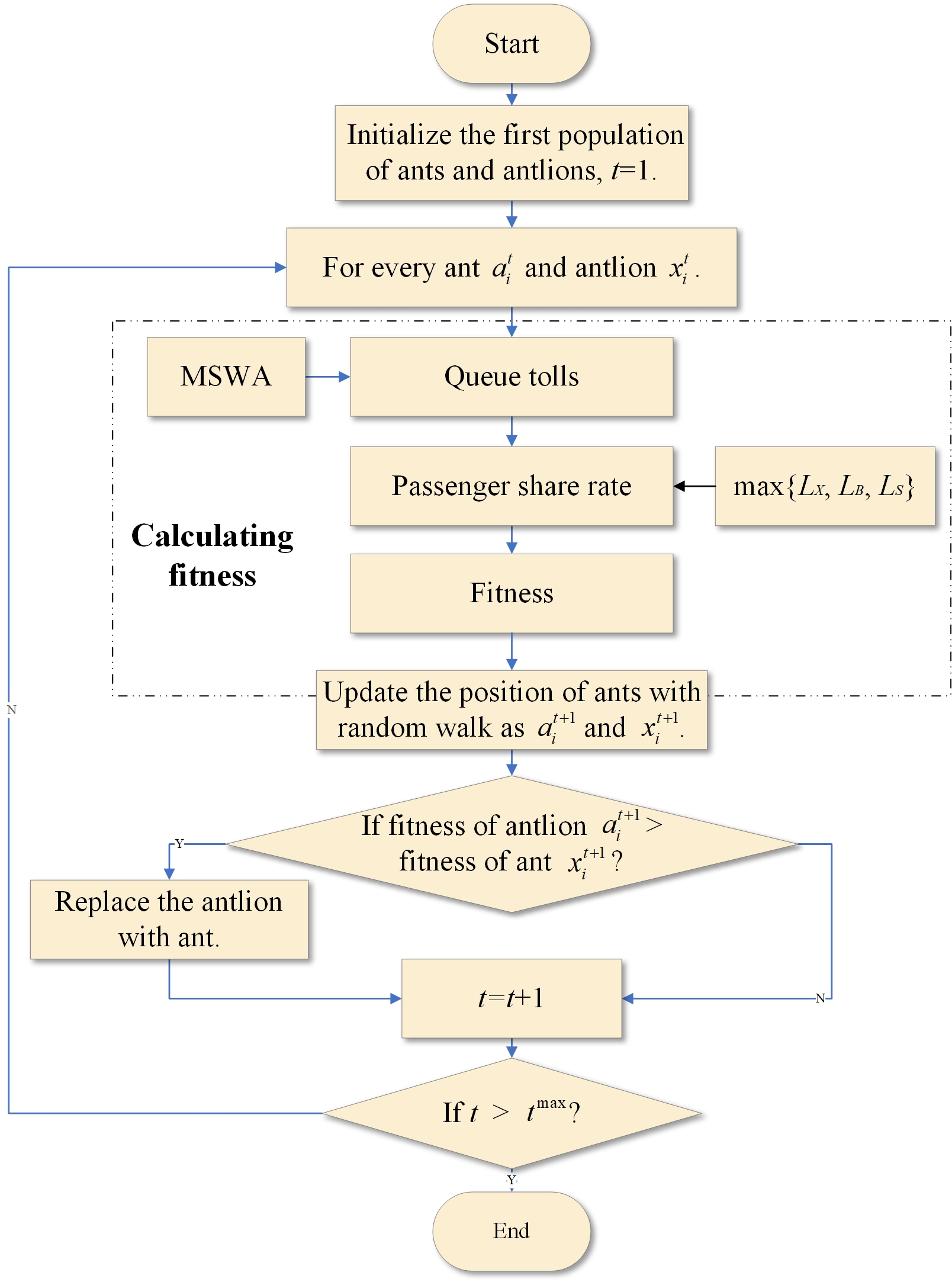}
\caption{Flow chart of solving bi-level programming model with ALO algorithm.}
\label{Hu5}
\end{figure}

The entire solution of bi-level programming is illustrated in Fig. \ref{Hu5}.

\section{Numerical simulation}\label{section6}

In this section, we simulate the proposed model in MATLAB 2020a. Section \ref{section6.1} predicts the queue length variation under different passenger arrival rates obtained by RK4, while Section \ref{section6.2} presents the solution procedure of passenger share rate employing MSWA by utilizing different groups of initial static utility. Section \ref{section6.3} presents and analyzes the results of solving bi-level programming model under two congestion scenarios.

In this work, the queueing network of airport ground public transport system is just a basic model considering most airport situations, and the real precise parameters in airport access queueing network are all different. The specific data acquisition is not involved in this paper, hence the parameters for numerical simulation are crude estimations.

\subsection{Instantaneous state of the M/M/c/K model solved by RK4}\label{section6.1}

\begin{figure*}
\centering
		\begin{subfigure}{.45\textwidth}
			\centering
			\includegraphics[width=\textwidth]{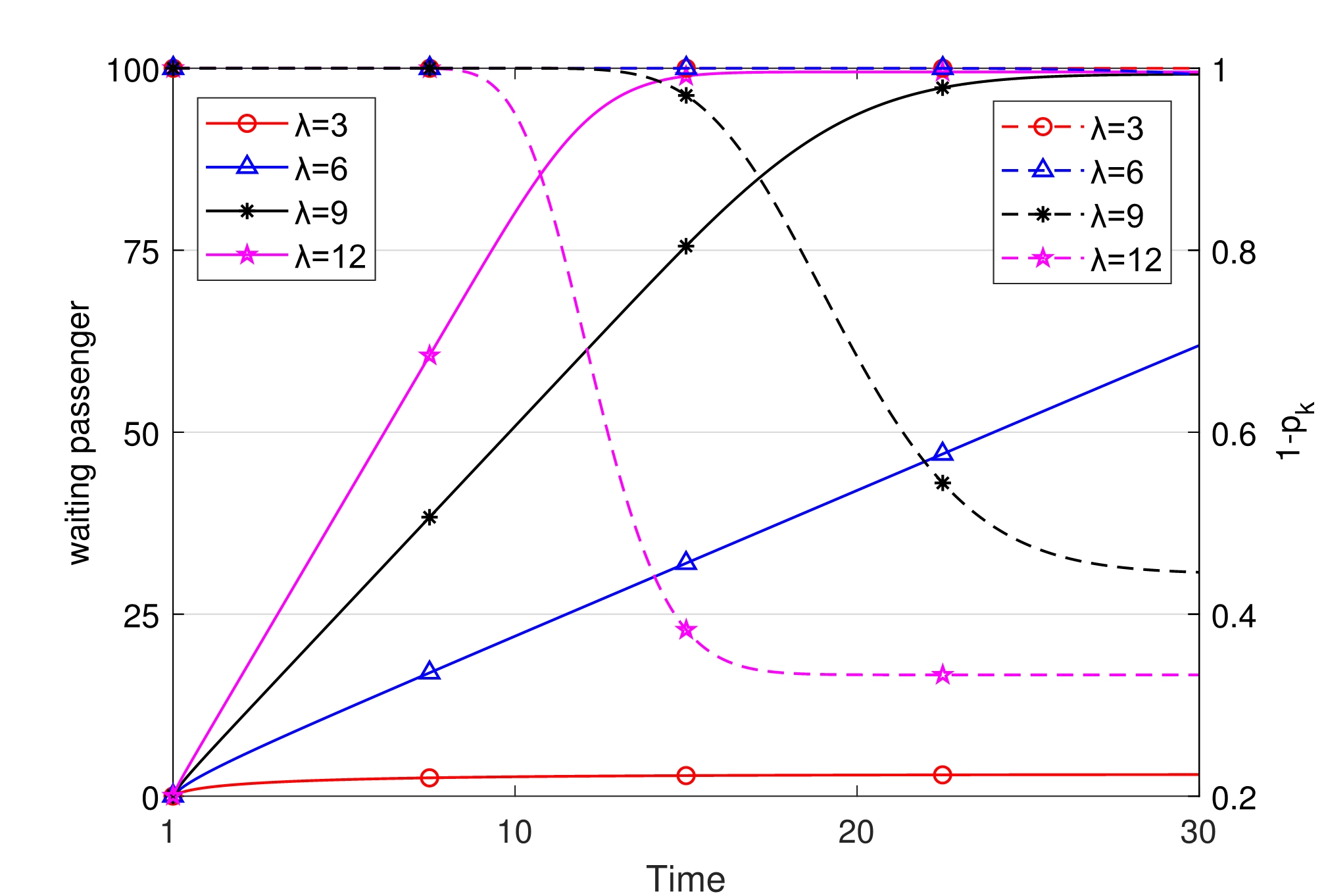}
            \label{Hu6a}
			\caption{Predicted queue length within 30-time units with $\lambda$ = 3,6,9,12 of queueing system.}
		\end{subfigure}
		\begin{subfigure}{.45\textwidth}
			\centering
			\includegraphics[width=\textwidth]{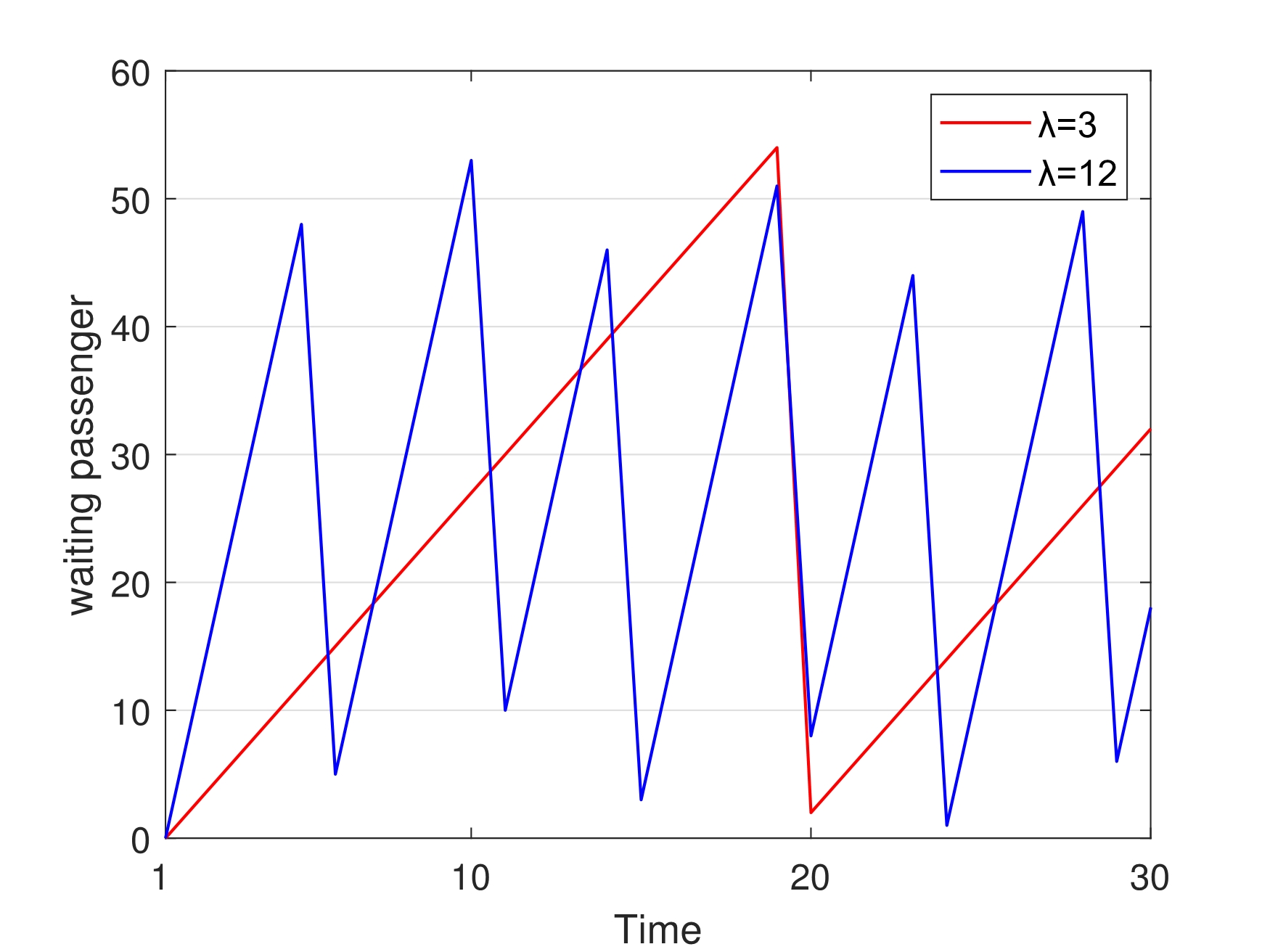}
            \label{Hu6b}
			\caption{Predicted queue length within 30-time units with $\lambda$ = 3,12 of Min(N,T) rule.}
		\end{subfigure}
        \caption{Predicted queue length of queueing system and Min(N,T) renewal process.}
        \label{Hu6}
\end{figure*}

This paper employs the queuing theory models and Min(N,T) renewal process to predict the queue length variations in different transport systems. We exploits a certain M/M/c/K model with parameters $\mu=0.5, c=8, K=100$, $P(0)=[1,0,0,\dots,0]$, and $t=0.005$, and a Min(N,T) renewal process with parameters $N=55$, $T=40$, to simulate their state variance using RK4 method and Eq. \ref{eq17}. Depending on whether the system utilization $\rho$ is less than 1, the queueing system is steady or not. For the trials four-passenger arrival rates are employed, $\lambda=3,6, 9, 12$. For $\lambda=3$,$\rho=0.75<1$ and the queue length tend to stabilize over time. For $\lambda = 6$, $\rho=1.5,2.25,3{\ } ( \geq1)$, and the queue length continuously increases. The expected instantaneous queue lengths within 30-time units are calculated with the results illustrated in Fig. \ref{Hu6}.

In the simulation results of M/M/c queueing system, for $\lambda=3$, the queue length increases from the trial start till the $15^{th}$ time unit. For $\lambda=6$, the queue length continuously increase without reaching the waiting area size $K$, while for $\lambda=9,12$, the queue lengths increase pretty fast, reaching the passenger waiting area size in 30-time units. We conclude that the faster the passenger arrival rate is, the faster the queue length will grow, and the more severe the passenger congestion will be. These findings are consistent with the theory of stable distribution in a queueing model.

Fig. \ref{Hu6}a indicates that the greater the initial passenger arrival rate $\lambda$, the greater the probability of passenger loss. Moreover, as time passes, the probability of passenger loss continues to increase, and the expected queue length with an unstable condition will be closer to the passenger waiting area size $K$. According to Fig. \ref{Hu6}a, for $\lambda=3,6$, the passenger loss probabilities are low and can be ignored. For $\lambda=9$ and $12$, the passenger loss probabilities exceed 0.5 and 0.6 as waiting area being full, and $p_{K}$ is not approximately equal to 0 as the queue length exceeds three-fourths of the waiting area size $K$. When the passenger waiting area is full, the system will not accept new arriving passengers, i.e., passenger loss case. For the ground access system examined in this paper, airport should ensure that the passenger waiting areas meet passenger demands, thus passenger waiting areas are assumed unlimited, while an integer is required to approximate the room of passenger waiting area for numerical simulation when passenger amount exceeds it the corresponding probability can be negligible. The results of Fig. \ref{Hu6}a could be helpful on choosing an appropriate integer for simulation.

Fig. \ref{Hu6}b shows the expected passenger number variance of Min(N,T) renewal process. Compared with queueing theory model, the expected passenger numbers of Min(N,T) renewal process change differently under the same passenger arrival rate. The passenger number of queueing system and Min(N,T) renewal process also have different sensitivity with passenger arrival rate and time. This difference between queueing model and Min(N,T) rule would further influence the solution results of lower level function and bi-level programming model using MSWA and ALO algorithm.
\subsection{MSWA for passenger share rate optimization}\label{section6.2}

For the given static utility in lower level subprogram, MSWA method iteratively estimates the waiting time of three transport modes and the passenger share rate among taxi, bus and subway with the known initial state of queueing network. To demonstrate the effectiveness of MSWA in calculating the passenger share rates, we set the various transport static utility values shown below and explore their impacts on passenger queueing time.

\begin{description}
\item[Case 1] $O_{X1}=0.5,O_{X2}=0.2,O_{B1}=1,O_{B2}=0.8,O_{S2}=0.3,O_{S2}=0.2$,
\item[Case 2] $O_{X1}=0.3,O_{X2}=0.2,O_{B1}=0.4,O_{B2}=0.5,O_{S2}=1,O_{S2}=0.8$.
\end{description}

\begin{figure*}
\centering
		\begin{subfigure}{.45\textwidth}
			\centering
			\includegraphics[width=\textwidth]{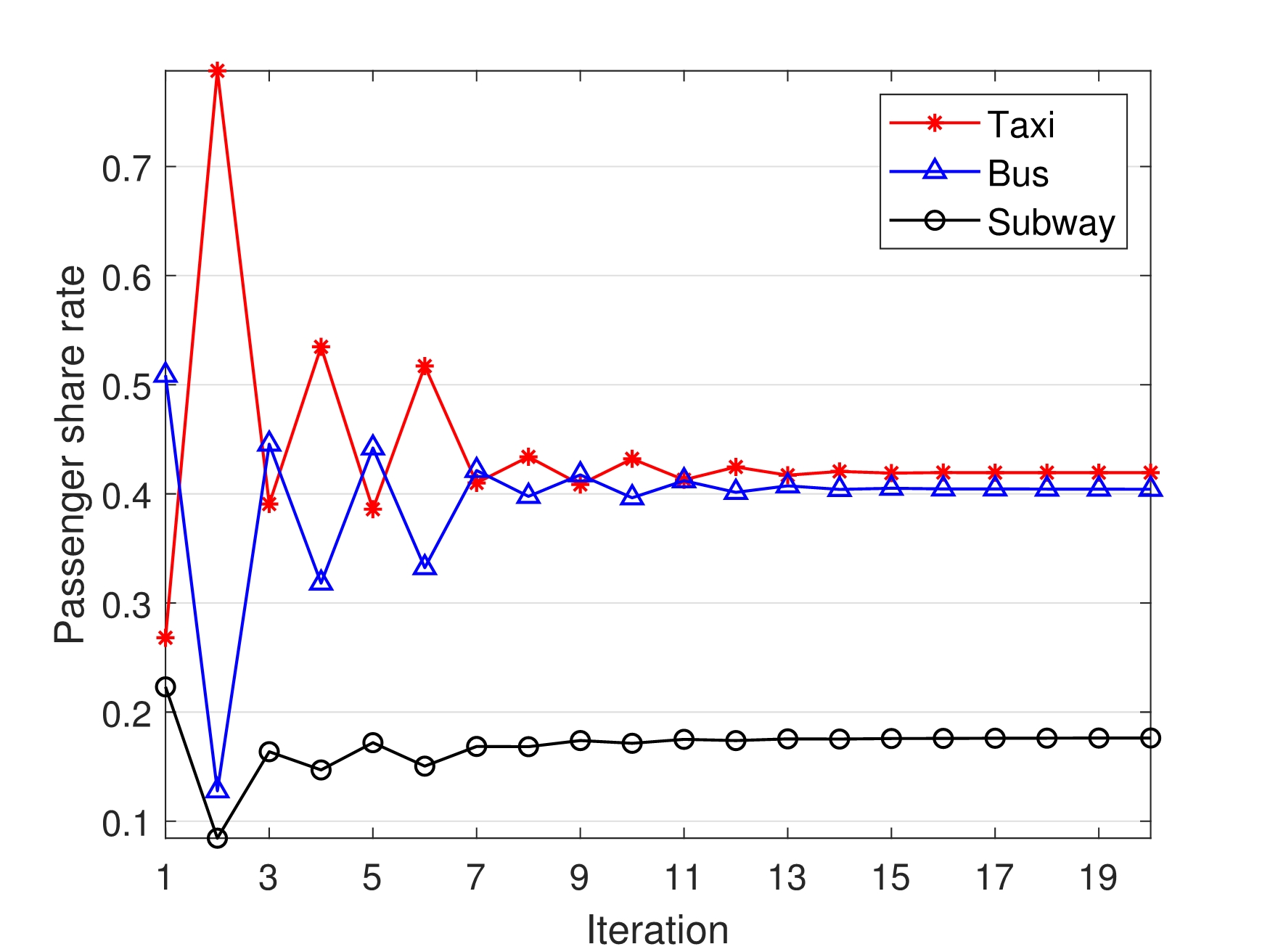}
            \label{Hu7a}
			\caption{Case 1}
		\end{subfigure}
		\begin{subfigure}{.45\textwidth}
			\centering
			\includegraphics[width=\textwidth]{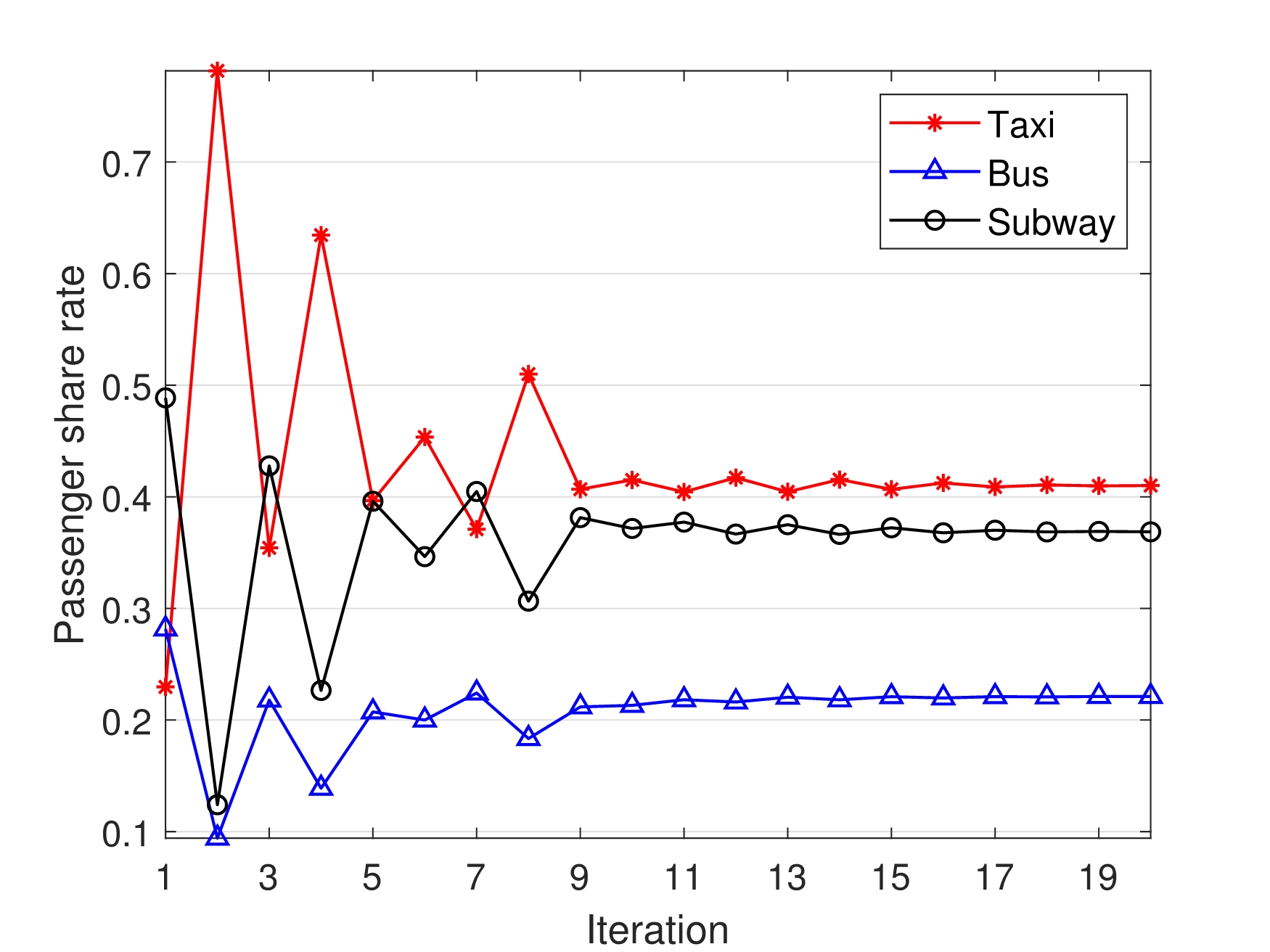}
            \label{Hu7b}
			\caption{Case 2}
		\end{subfigure}
        \caption{Passenger share rate histories of different cases.}
        \label{Hu7}
	\end{figure*}

\begin{figure*}
\centering
		\begin{subfigure}{.45\textwidth}
			\centering
			\includegraphics[width=\textwidth]{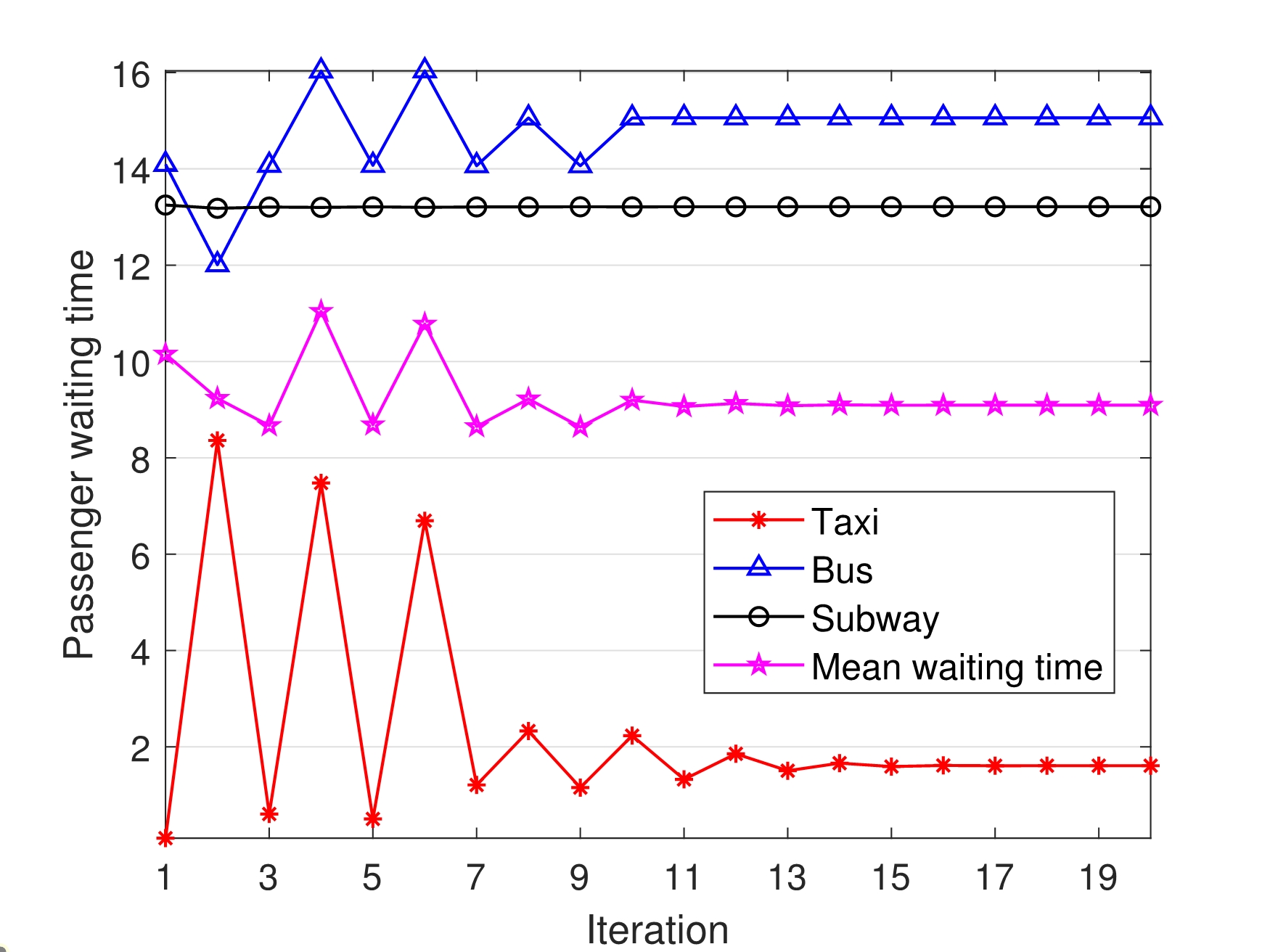}
            \label{Hu8a}
			\caption{Case 1}
		\end{subfigure}
		\begin{subfigure}{.45\textwidth}
			\centering
			\includegraphics[width=\textwidth]{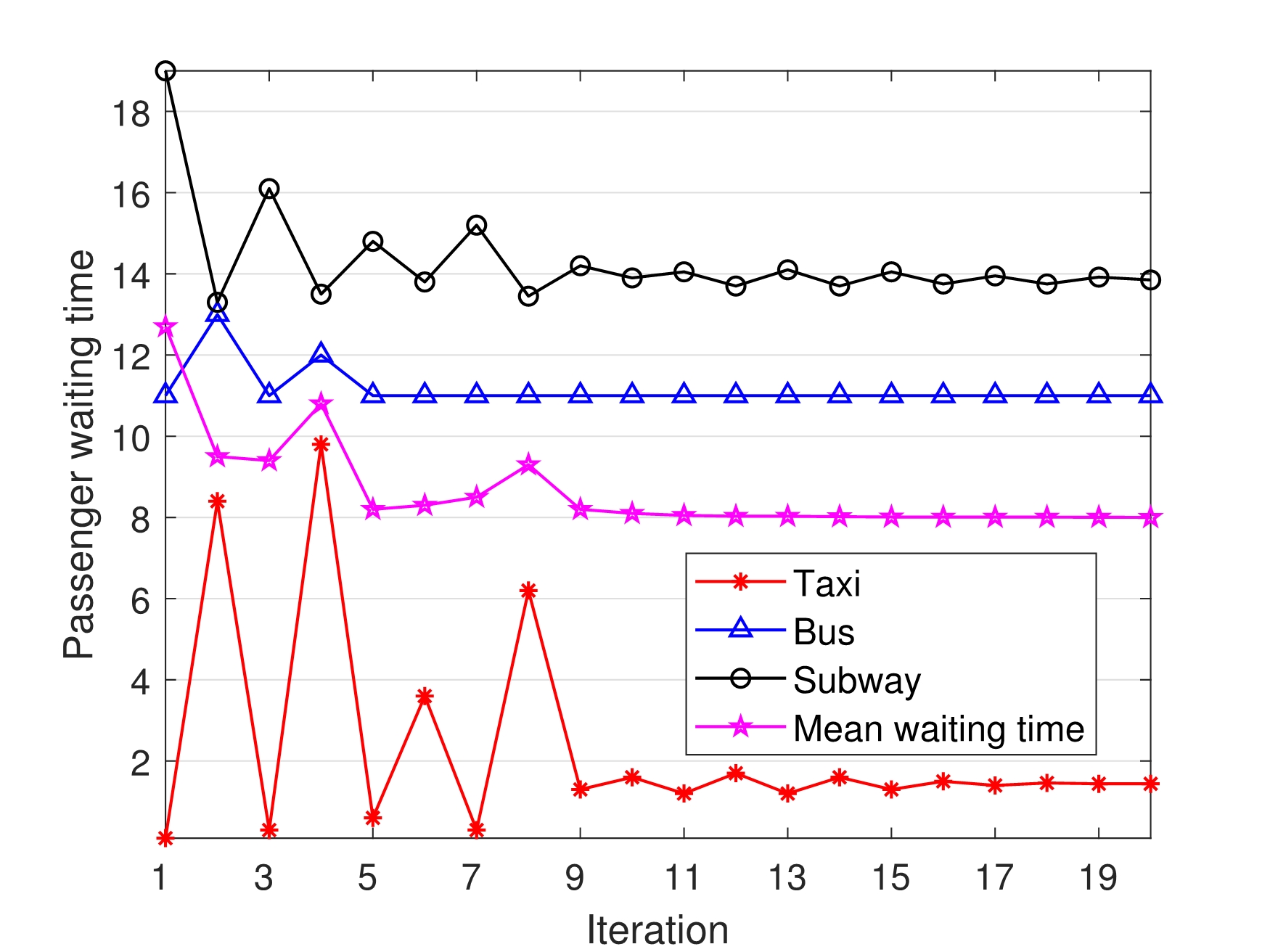}
            \label{Hu8b}
			\caption{Case 2}
		\end{subfigure}
        \caption{Expected passenger waiting time histories of different cases.}
        \label{Hu8}
	\end{figure*}

\begin{figure*}
\centering
		\begin{subfigure}{.45\textwidth}
			\centering
			\includegraphics[width=\textwidth]{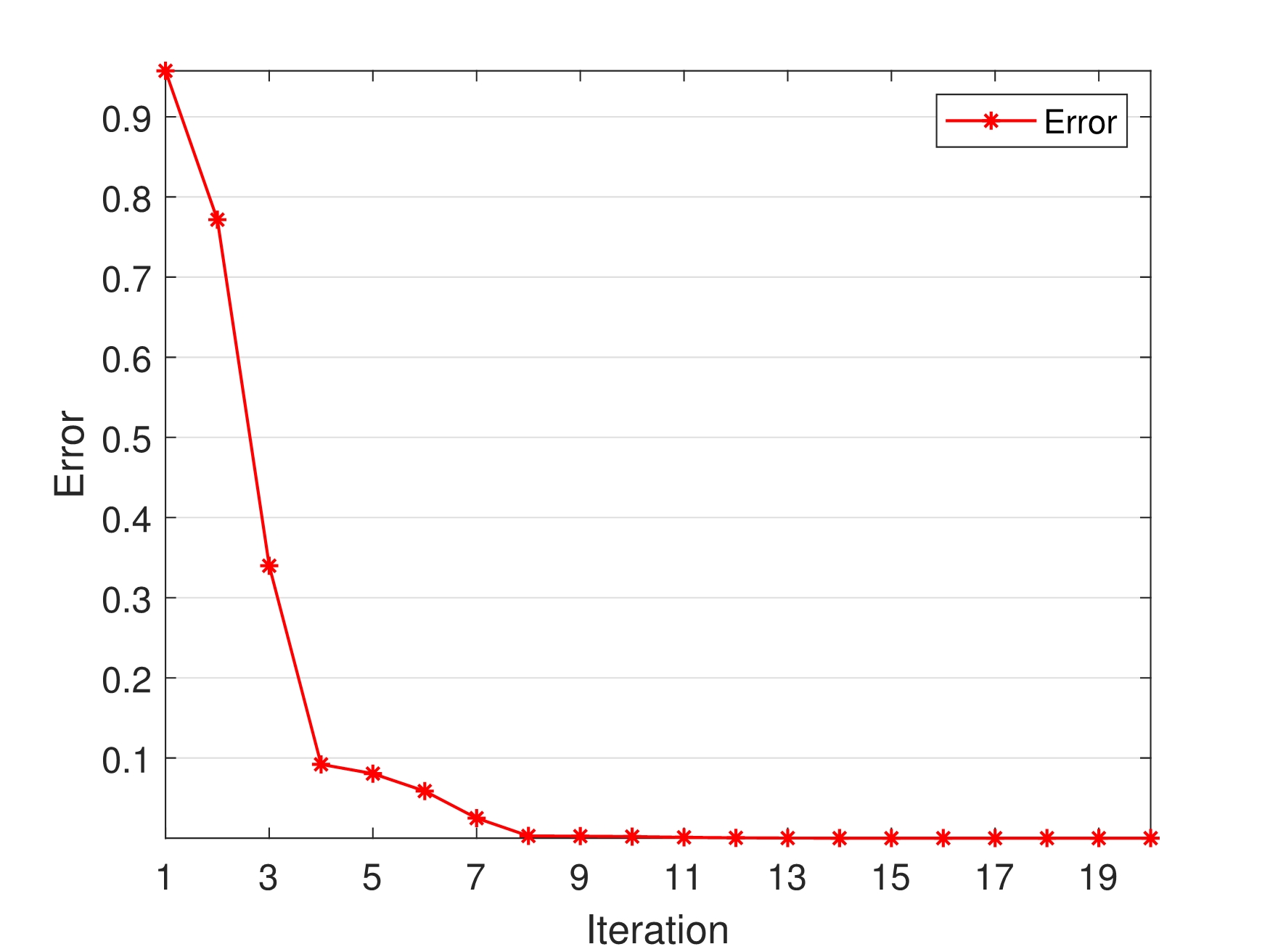}
            \label{Hu9a}
			\caption{Case 1}
		\end{subfigure}
		\begin{subfigure}{.45\textwidth}
			\centering
			\includegraphics[width=\textwidth]{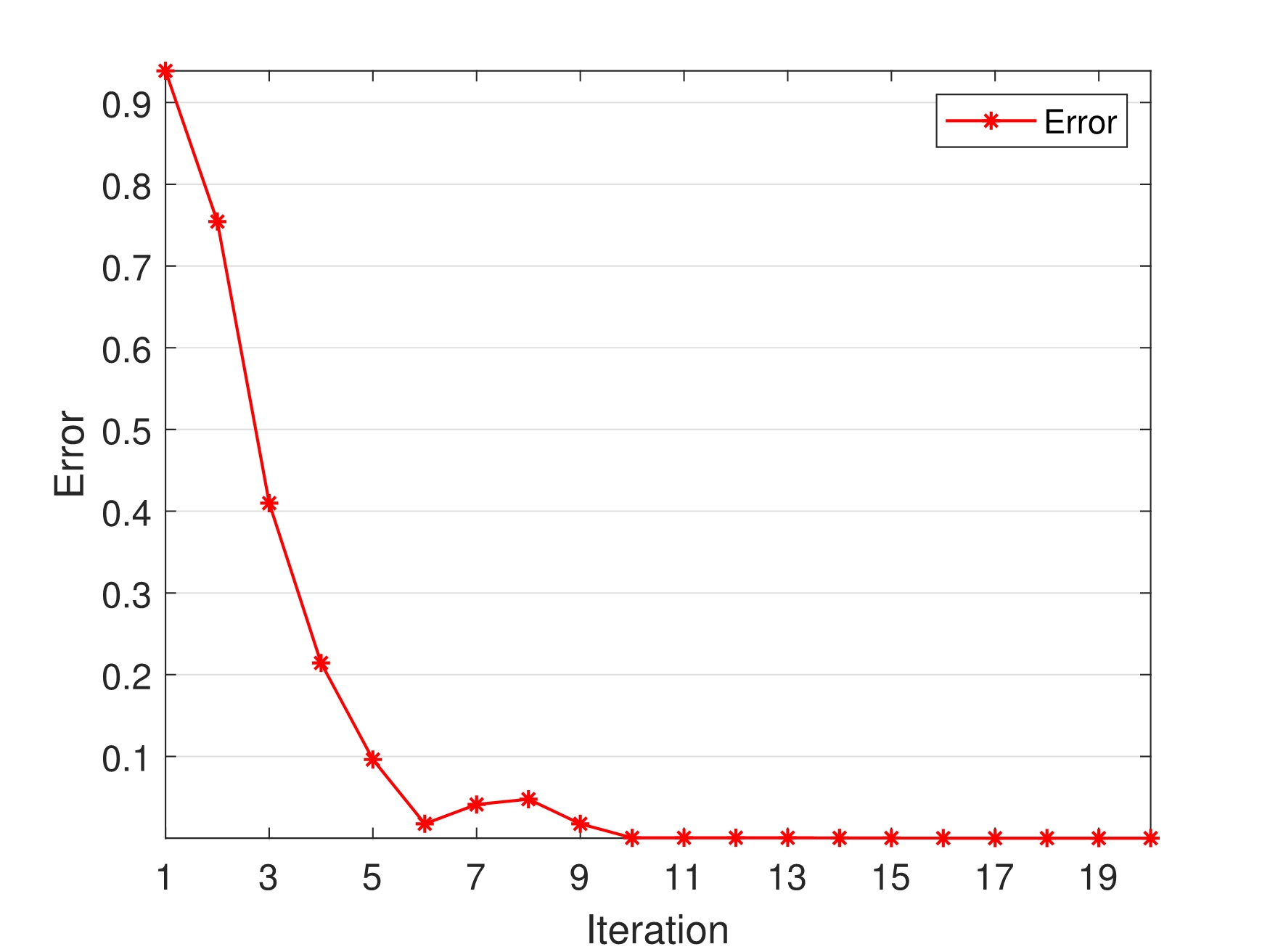}
            \label{Hu9b}
			\caption{Case 2}
		\end{subfigure}
        \caption{Error histories of different cases.}
        \label{Hu9}
	\end{figure*}

In this trial, we choose two classes of passengers with different static utilities shown above, and set the weights of the static utilities and the queueing time as 1:1 for all classes of passengers. The corresponding queueing time, passenger share rates, and convergence errors are illustrated in Figs. \ref{Hu7}-\ref{Hu9}.

In Fig. \ref{Hu7}, the passenger share rates initially fluctuate around the final convergence value, which reflect that, the growth of passenger share rates in taxi and subway would cause the same increase on corresponding waiting time which would negatively feed back on passenger share rate. After several iterations, all share rates tend to converge to the specific values, and the passenger flow distribution reaches an equilibrium under the increment-iteration by MSWA. Passenger share rates are initially determined by the static utilities but finally affected by transport service rate. This equilibrium process of passenger share rate and waiting time actually reduce the imbalance between passenger flow demand and service rate supply in one transport, and make the differences between single transport's demand and supply in taxi, bus and subway get closer. The solution of MSWA is highly correlated with waiting time's sensitivity to passenger arrival rate, thus the mean waiting time can always decline after airport provide the waiting time estimation to passengers.

 Fig. \ref{Hu8} indicates that the passenger average queueing time has a downtrend as the iteration process evolves, hence the queue congestion is alleviated to some extent. The variances of waiting time with iterations in taxi and subway are consistent with corresponding share rate, but not necessarily in bus as shown in Fig. \ref{Hu8}a, which is due to the Min(N,T) departure rule of bus that the more passenger arrival may lead the bus getting full earlier and departing immediately. Despite this characteristic, the Min(N,T) departure rule is less sensitive to passenger arrival rate than the queueing system of taxi and subway, passenger flow do not shift towards bus during MSWA iterations. In order to balance the passenger flows among transport modes, the decrease of the waiting time in one transport mode must be at the cost of increasing the waiting time in another transport mode, still the total average passenger waiting time is reduced, meeting the decision maker's requirement.

Fig. \ref{Hu9} illustrates the iterative errors that decrease within the given limit after several iterations, inferring that the passenger share rate optimization method is indeed feasible and does converge. Therefore, it is concluded that the MSWA affords to reduce the mean queueing time by optimizing the passenger share rates.

\subsection{ALO for generating queue toll scheme}\label{section6.3}

\begin{figure}[h]
\centering
		\begin{subfigure}{.45\textwidth}
			\centering
			\includegraphics[width=\textwidth]{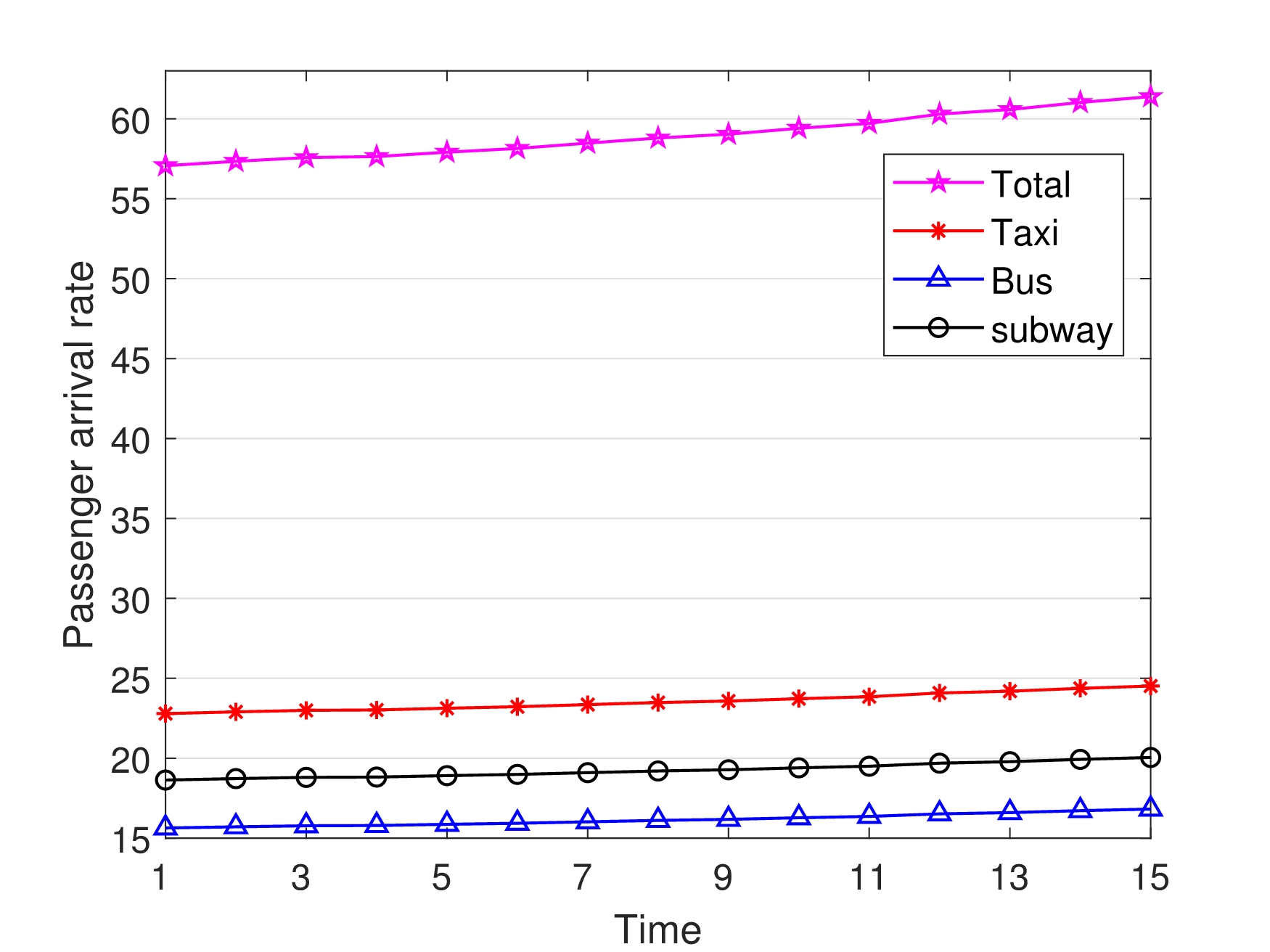}
            \label{Hu10a}
			\caption{Passenger arrival rate within 15 minutes in the daytime.}
		\end{subfigure}
		\begin{subfigure}{.45\textwidth}
			\centering
			\includegraphics[width=\textwidth]{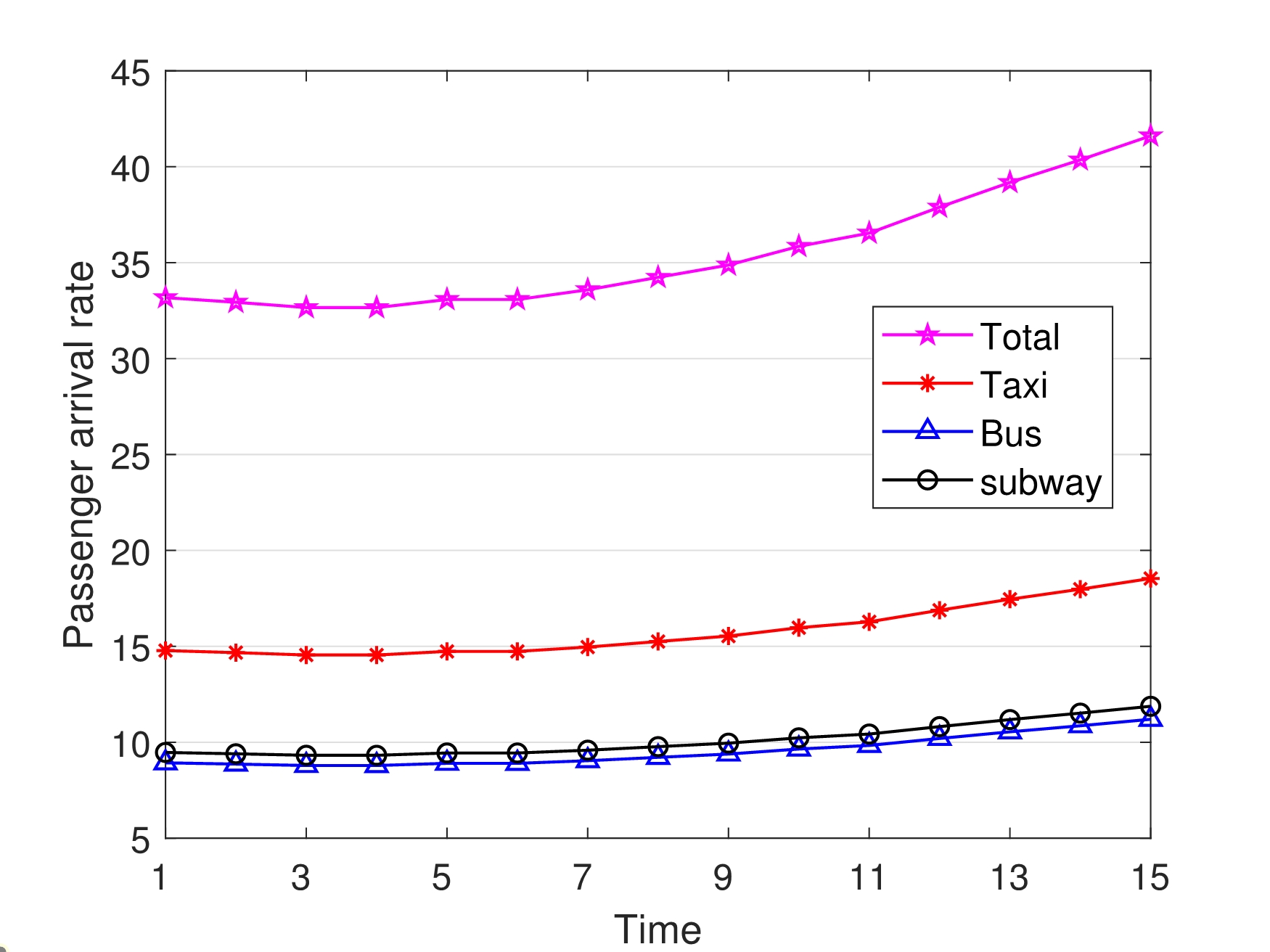}
            \label{Hu10b}
			\caption{Passenger arrival rate within 15 minutes in the evening.}
		\end{subfigure}
        \caption{Passenger arrival rate under two scenarios of simulation.}
\label{Hu10}
\end{figure}

The airport ground public transport queue congestions always appear in two time periods of a day, for early afternoon and late evening, with different causes and characteristics. At around 2 p.m., the passenger flow reaches the peak of a day, queue time for taking taxi and subway grows rapidly because of the imbalance between passenger flow and service capacity, while that of bus queueing system may decrease instead because of its Min(N,T) departure rule. At about 10 p.m., the hour that airport bus and subway shut soon, the total passenger flow partly declines compared with the daytime. Passengers would hold a stronger intention to take taxi (people really want to hurry home as getting late), thus taxi system shares more passengers than afternoon, while its taxi supply decreased even further at evening. Bus departure interval also increase at this time period. When the predicted number of stranded passenger exceed an acceptable range, decision makers start a queue toll program. In this section, we simulate the queue toll programs based on the proposed bi-level programming model under the two scenarios above. The initial conditions, queueing network parameters and arrival rate are illustrated in Table. \ref{Hut1} and Fig. \ref{Hu10}. For the taxi arrival simulation data, its arrival rate is set according to flight departure timetable, based on the inference that taxis taking the passenger to airport always want to carry a passenger and leave the airport.

\begin{figure*}
\centering
		\begin{subfigure}{.8\textwidth}
			\centering
			\includegraphics[width=\textwidth]{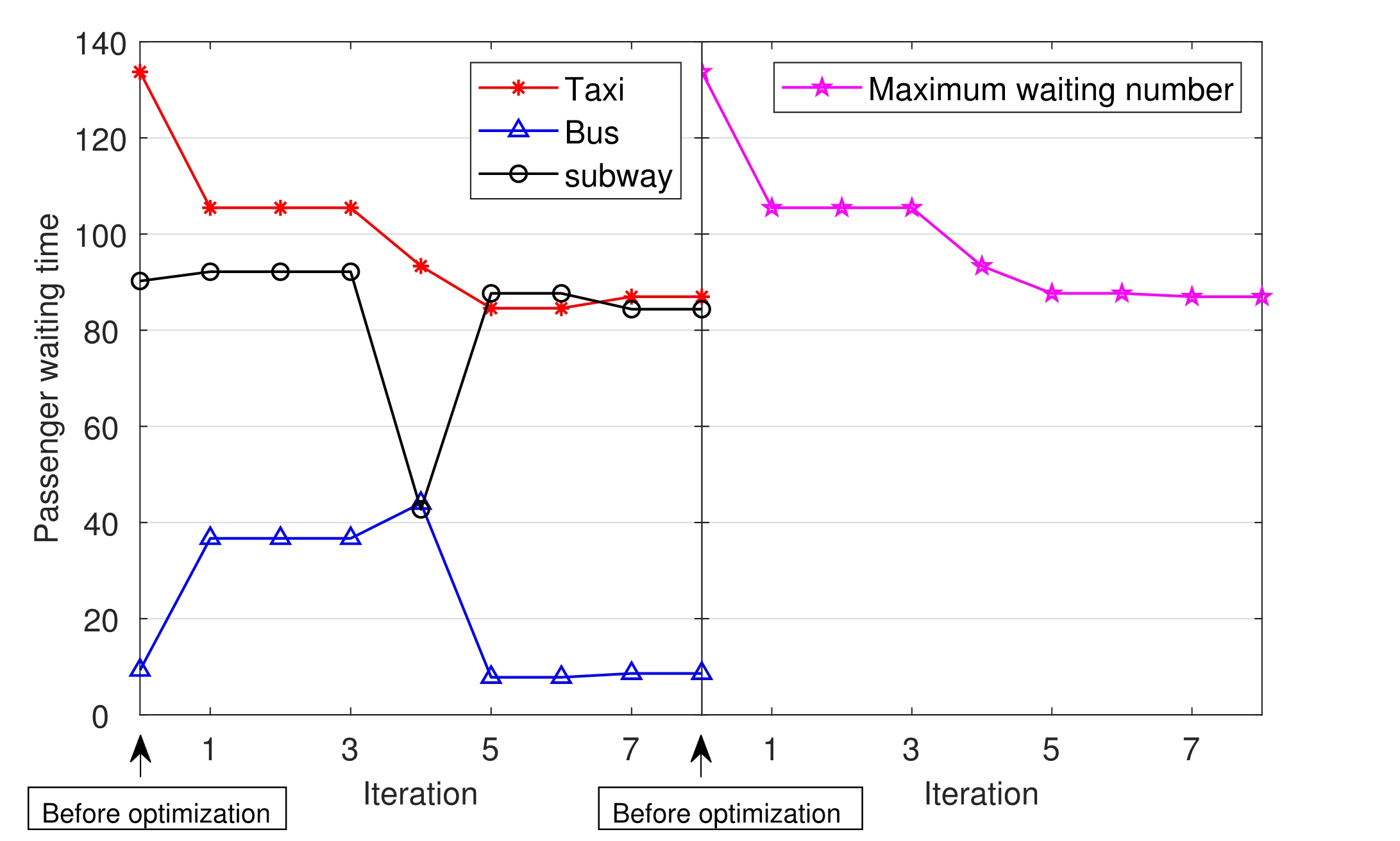}
            \label{hu11a}
			\caption{The number of stranded passengers in each transport mode, and their maximum value during iterations. }
		\end{subfigure}
		\begin{subfigure}{.45\textwidth}
			\centering
			\includegraphics[width=\textwidth]{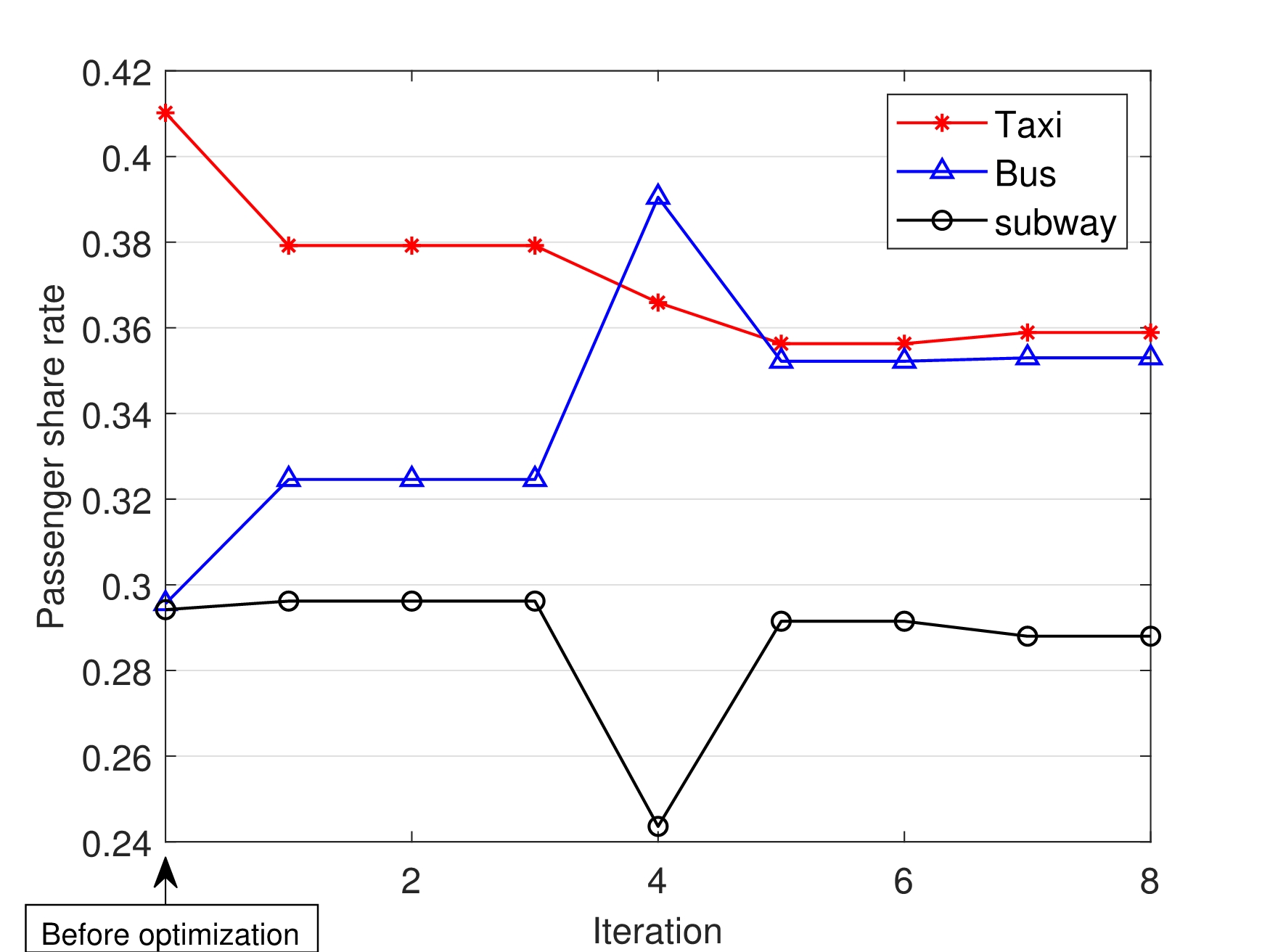}
            \label{hu11b}
			\caption{The variance of passenger share rate with iterations.}
		\end{subfigure}
		\begin{subfigure}{.45\textwidth}
			\centering
			\includegraphics[width=\textwidth]{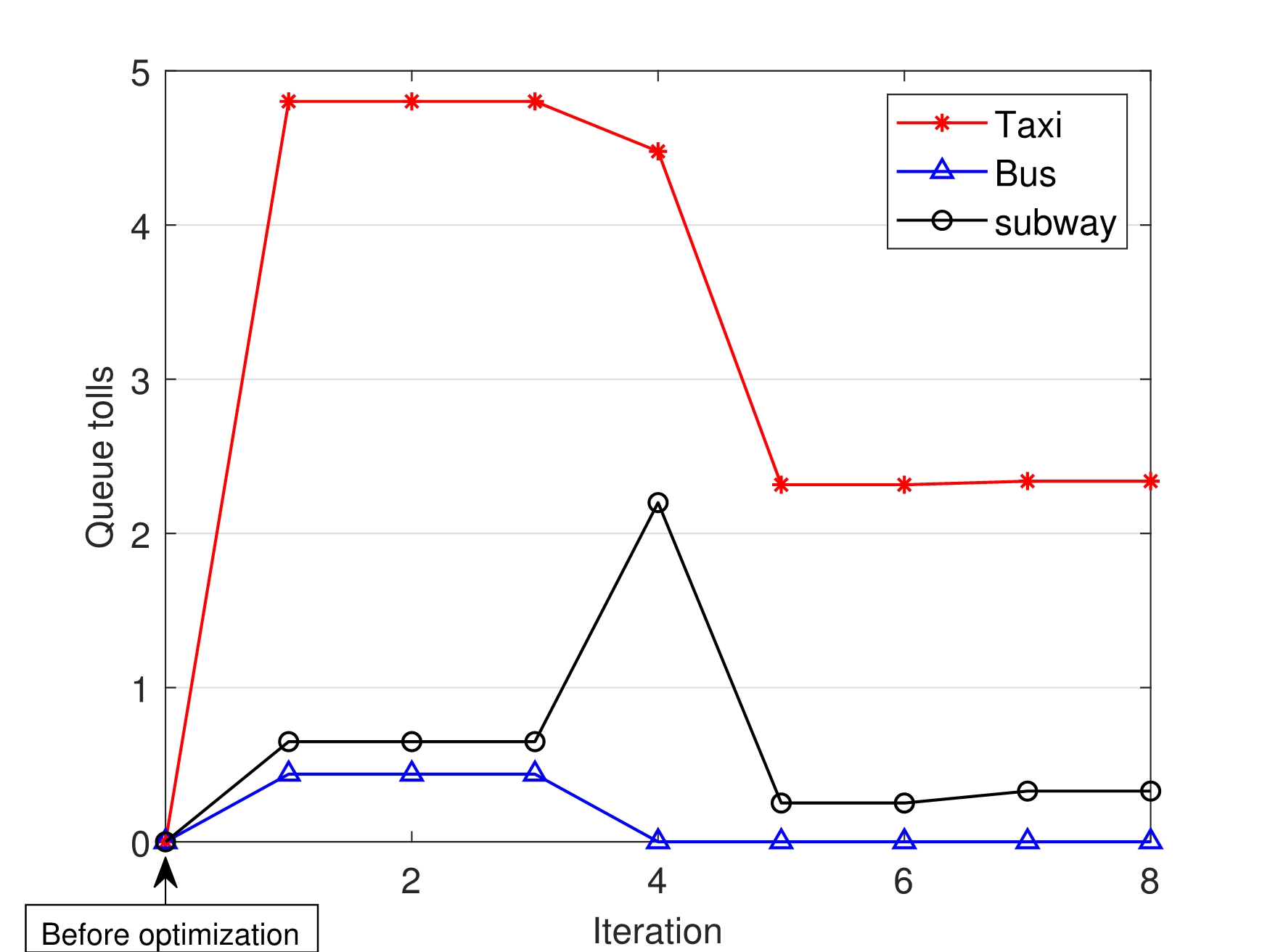}
            \label{hu11c}
			\caption{The best queue toll scheme of every step of optimization.}
		\end{subfigure}
        \caption{Results of simulating bi-level programming model under the daytime condition.}
        \label{hu11}
	\end{figure*}

In the trial of the mass passenger flow condition by day, at the initial time, the current number of passenger staying in taxi bus and subway systems are 60, 8 and 63 respectively. The prediction of queueing network state shows that, after 15 minutes of service with the passenger flow in Fig. \ref{Hu10}a, the number of stranded passengers will further reach 137, 6 and 84 before optimization, which will exceed airport's affordability. We carry out 8 steps of ALO iterations to search the queue toll scheme for minimizing Eq. \ref{eq29}, and the variance of stranded passenger numbers, passenger share rate and elite antlion (best queue toll scheme) during the iteration process is shown in Fig. \ref{hu11}.

In Fig. \ref{hu11}c, the queue tolls for taxi, bus and subway in the first search are 4.8, 0.4 and 0.6 respectively; after several steps, the elite antlion converges to queue tolls of 2.3, 0 and 0.3. In Fig. \ref{hu11}b, the queue toll scheme makes passengers who originally planned to take the taxi and subway shift towards airport bus, thus the stranded crowd in taxi system decline significantly. The airport bus shares more passengers after imposing the converged queue toll scheme, while its number of stranded passengers almost unchanged owing to the Min(N,T) departure rule. Observing passenger share rate and maximum passenger number in Fig. \ref{hu11}b and Fig. \ref{hu11}a, the passenger wait and bus departure process during the 15 minutes can be inferred. In the no queue toll situation and first three iterations, the numbers of stranded passengers reduce at the fifteenth minute compared with initial condition, indicating that there is a bus depart advance in this period resulting from being full advance, and the passengers getting on the next bus accumulated to 6 and 39 at the fifteenth minute. In the last 4 iterations, there are two buses depart advance. This suggests that airport bus is a great choice to transfer passenger flow pressure of taxi and subway.

\begin{figure*}
\centering
		\begin{subfigure}{.8\textwidth}
			\centering
			\includegraphics[width=\textwidth]{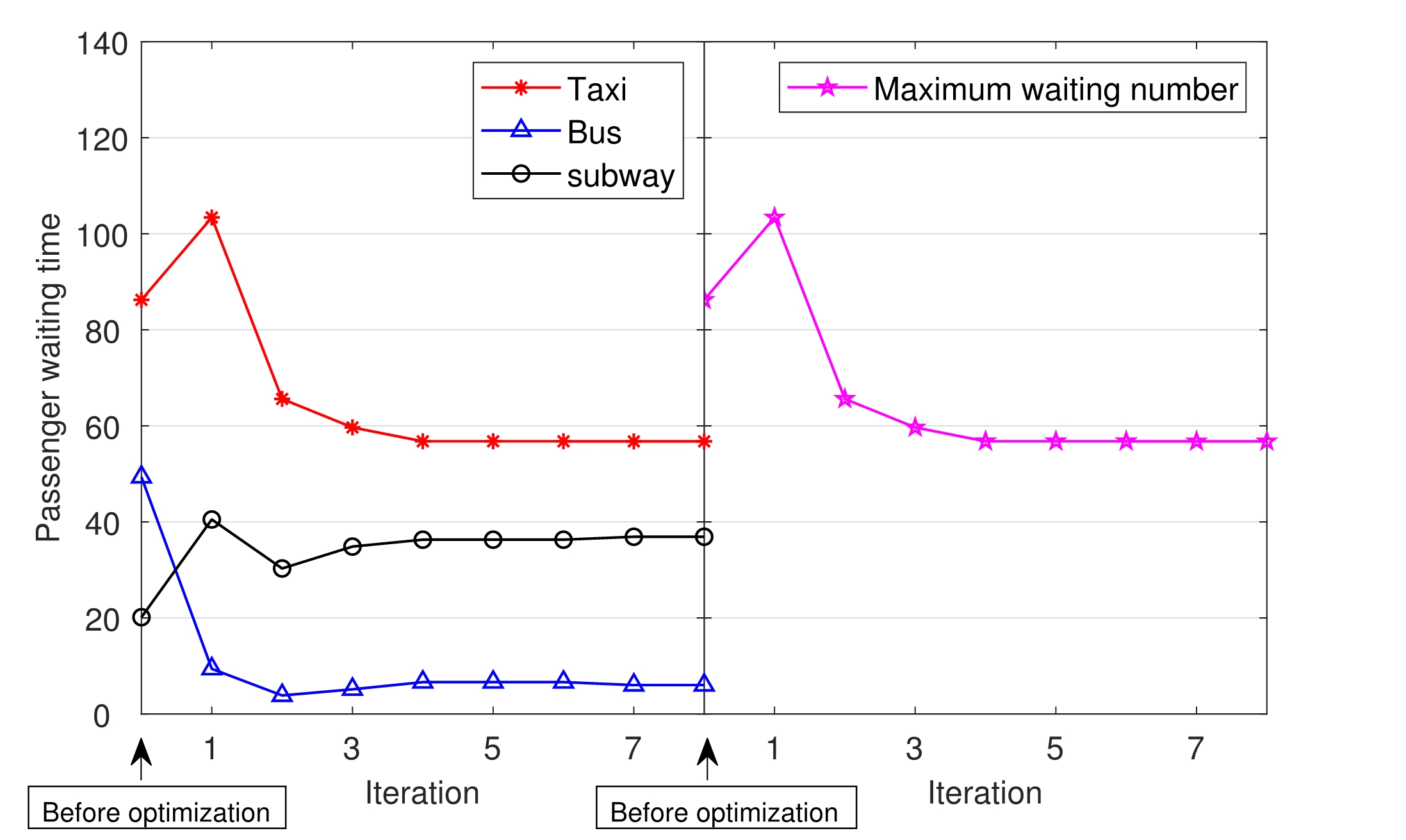}
            \label{hu12a}
			\caption{The number of stranded passengers in each transport mode, and their maximum value during iterations. }
		\end{subfigure}
		\begin{subfigure}{.45\textwidth}
			\centering
			\includegraphics[width=\textwidth]{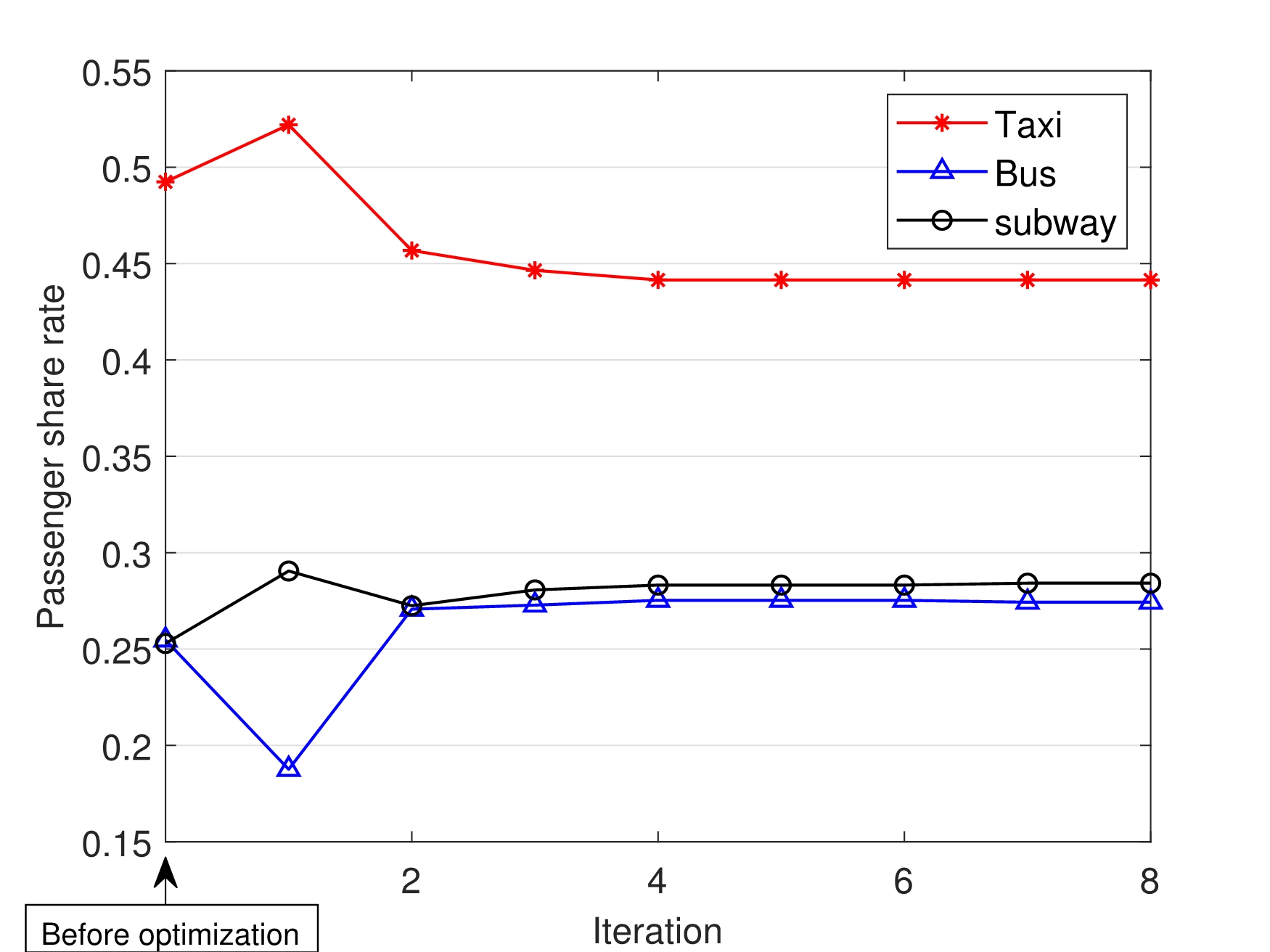}
            \label{hu12b}
			\caption{The variance of passenger share rate with iterations.}
		\end{subfigure}
		\begin{subfigure}{.45\textwidth}
			\centering
			\includegraphics[width=\textwidth]{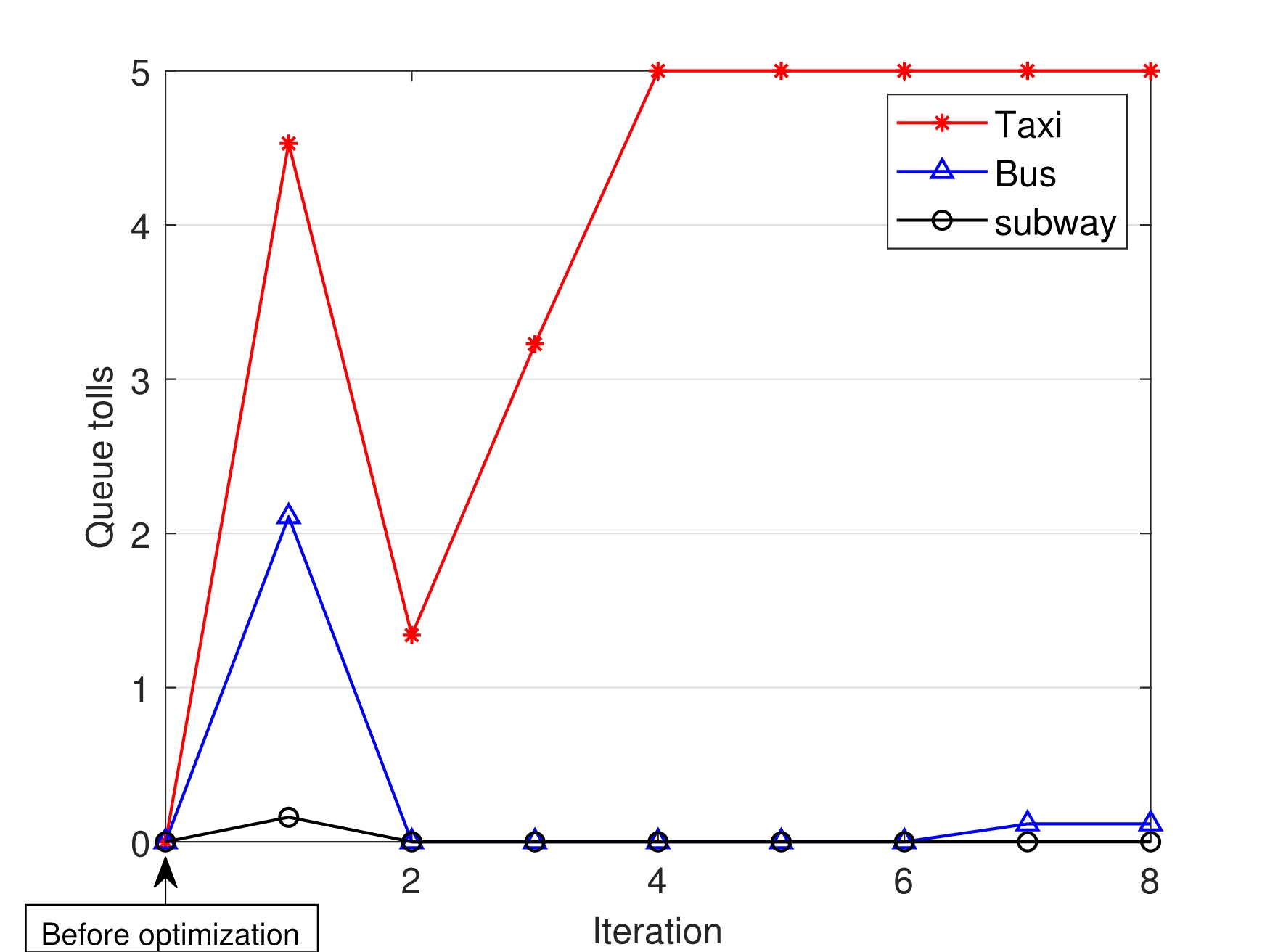}
            \label{hu12c}
			\caption{The best queue toll scheme of every step of optimization.}
		\end{subfigure}
        \caption{Results of simulating bi-level programming model under the evening condition.}
        \label{hu12}
	\end{figure*}

In the second trial under the evening condition, the number of passengers staying in taxi, bus and subway system are 33, 7 and 21 respectively. After 15 minutes, stranded passenger number will reach services with passenger flow of Fig. \ref{Hu10}b, the number of stranded passengers will further reach 82, 50 and 20 by forecast. The results of 8 steps optimization by ALO are illustrated in Fig. \ref{hu12}.

In Fig. \ref{hu12}c, the queue tolls for taxi, bus and subway finally converge to queue tolls of 5, 0.1 and 1. Fig. \ref{hu12}a shows that in the evening, queue congestion mainly appears in taxi queueing system, in which the numbers of stranded passengers are always the maximum during iterations. In Fig. \ref{hu12}b, bus and subway share part of potential passengers of taxi under the best queue toll scheme, thus taxi queue length reduce. Compared with the daytime, airport bus can not share as much passenger flows as it in the evening. With less passenger arrival rate, waiting time for bus becomes insensitive to the variance of passenger flow, while the rate of change of taxi and subway queue time with respect to passenger arrival rate is still high. Hence subway shares most potential heavy passenger traffic of taxi after imposing queue tolls in the evening.

Under the queue toll program, the redistribution of potential passenger flow among three transport modes is jointly completed by MSWA and ALO. The MSWA is associated with the sensitivity between queueing time and passenger arrival, and ALO algorithm minimizes the maximum stranded passenger number of three transport modes, which all balance the passenger arrival in three transport modes.

\section{Conclusion}\label{section7}
This paper proposes a bi-level programming model for airport ground access system congestion alleviation, in which the lower level solves the passenger choice problem based on MNL model, and the upper level subprogram optimizes queue length by imposing queue tolls on passengers to adjust their own choice. We exploit a queueing network model considering most airport situations to simulate the basic ground-access service process in airports, in which the queueing and service of three common airport ground access subsystems are further modeled. The developed model affords to predict the queue length according to the current network state and estimated passenger arrival, and adequately reflect the features of waiting process in airport ground access system. In the simulation, the developed MSWA method can iteratively solve the waiting time and passenger share rate in the lower level programming and make the passenger share rate converge to its equilibrium solution with waiting time. Finally, we implement the bi-level programming to generate queue toll scheme and impose it on passenger. Numerical experiments considering the cases of daytime and evening, investigate these congestion characteristics and the capability of the proposed queue toll program in alleviating queue congestion in the airport ground-access system.

Despite the proposed model and algorithm attain an appealing performance, further improvements of the airport queue toll program are still required. In this work, the queueing time prediction relies on the predicted passenger arrival rate based on flight timetable, and its forecast accuracy has room for growth. Hence, future work should consider the combination of passenger flow and the queueing time predictions for even higher precision. Additionally, limited by available resources, we do not involve any precise practical data and field tests in this work. In future work, the effectiveness and advantages of the proposed model could be confirmed via practical tests at the airport.

\section*{Acknowledgement}
The work in this study was supported by the National Key Research and Development Program of China (grant no. 2018YFB1601200).

\bibliographystyle{tfcad}
\bibliography{Passenger_Congestion_Alleviation}

\newpage

\begin{table*}[!t]
\caption{Simulation parameters}
\label{Hut1}
\centering
\begin{tabular}{p{60 pt}p{50 pt}p{50 pt}p{140 pt}}
\hline
{Para}  &{Day} &{Night}& {Units}\\
\hline
$\lambda_{T}$&{0.05}&{0.05} &passenger of flight departure\\
$\mu_{X}$&{3} &{3}&passenger/minute\\
$\mu_{B}$ & {1}&{1} &passenger/minute\\
$\mu_{S1}$ & 8 &8 &passenger/minute\\
$\mu_{S2}$ & 1 & 1 &passenger/minute\\
$c_{B}$ &2&1&/\\
$c_{S1}$ & 2  & 1 &/\\
$c_{S2}$ & 2 &1&/\\
$N$ & {55}&55&passenger\\
$T$&{30}&{80}& minute\\
$O_{X}$& {1}&{1} &/\\
$O_{B}$ & {0.2}&{0.2}&/\\
$O_{S}$&{0.5}&{0.5}& /\\
$\omega^T_{X}$, $\omega^O_{X}$, $\omega^J_{X}$& {1 1.2 0.6}&{1 1.2 0.6} &/\\
$\omega^T_{B}$, $\omega^O_{B}$, $\omega^J_{B}$&{1 0.7 0.7}&{1 0.7 0.7}& /\\
$\omega^T_{S}$, $\omega^O_{S}$, $\omega^J_{S}$& {1 1  0.7}&{1.2 0.8 0.7} &/\\
\hline
{Initial state}  &{Day} &{Night}& {Units}\\
\hline
$L_{X}(t_0)$& 60&33 &passenger\\
$L_{B}(t_0)$&5&4& passenger\\
$N_0$&3&3& passenger\\
$T_0$&10&10& minute\\
$L_{S1}(t_0)$& 57&20 &passenger\\
$L_{S2}(t_0)$& 6&1 &passenger\\
$M_0$&2&2& minute\\
\end{tabular}
\end{table*}

\clearpage

\begin{figure}
\centering
\includegraphics[width=3.5in]{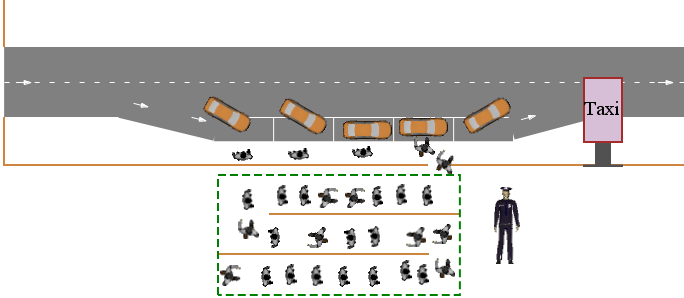}
\caption{Taxi service process.\label{Hu2}}
\end{figure}

\clearpage

\listoffigures

\end{document}